\documentclass[prx,reprint,superscriptaddress,onecolumn,amsmath]{revtex4-1}
\usepackage{url}
\usepackage{textcase}
\usepackage{graphicx}
\usepackage{amsfonts}
\usepackage{txfonts}
\usepackage{times}
\usepackage{subfig}
\usepackage{hyperref}
\usepackage{amsmath}
\usepackage[utf8]{inputenc}

\usepackage{float}

\usepackage{comment}

\usepackage{textcomp}

\usepackage{color}
\definecolor{gray}{gray}{0.8}

\usepackage[toc]{appendix}

\graphicspath{{./Figures/}} 

\newcommand{\argmin}{arg\,min}
\begin{document}
\title{The edge-disjoint path problem on random graphs by message-passing}

\author{Fabrizio Altarelli}
\email{Presently at Capital Fund Management, 23 Rue de l'Universit\'e, 75116 Paris, France}
\affiliation{DISAT and Center for Computational Sciences, Politecnico di Torino, Corso Duca degli Abruzzi 24, 10129 Torino, Italy}
\affiliation{Collegio Carlo Alberto, Via Real Collegio 30, 10024 Moncalieri, Italy}
\author{Alfredo Braunstein}
\affiliation{DISAT and Center for Computational Sciences, Politecnico di Torino, Corso Duca degli Abruzzi 24, 10129 Torino, Italy}
\affiliation{Human Genetics Foundation, Via Nizza 52, 10126 Torino, Italy }
\affiliation{Collegio Carlo Alberto, Via Real Collegio 30, 10024 Moncalieri, Italy}
\author{Luca Dall'Asta}
\affiliation{DISAT and Center for Computational Sciences, Politecnico di Torino, Corso Duca degli Abruzzi 24, 10129 Torino, Italy}
\affiliation{Collegio Carlo Alberto, Via Real Collegio 30, 10024 Moncalieri, Italy}
\author{Caterina De Bacco}
\email{caterina.de-bacco@lptms.u-psud.fr}
\affiliation{LPTMS, Centre National de la Recherche Scientifique et Universite Paris-Sud 11, 91405 Orsay Cedex, France.}
\author{Silvio Franz}
\affiliation{LPTMS, Centre National de la Recherche Scientifique et Universite Paris-Sud 11, 91405 Orsay Cedex, France.}

\begin{abstract}
We present a message-passing algorithm to solve the edge disjoint path problem (EDP) on graphs incorporating under a unique framework both traffic optimization and path length minimization. The \textit{min-sum} equations for this problem present an exponential computational cost in the number of paths. To overcome this obstacle we propose an efficient implementation by mapping the equations onto a weighted combinatorial matching problem over an auxiliary graph. We perform extensive numerical simulations on random graphs of various types to test the performance both in terms of path length minimization and maximization of the number of accommodated paths. In addition, we test the performance on benchmark instances on various graphs by comparison with state-of-the-art algorithms and results found in the literature. Our message-passing algorithm always outperforms the others in terms of the number of accommodated paths when considering non trivial instances (otherwise it gives the same trivial results). Remarkably, the largest improvement in performance with respect to the other methods employed is found in the case of benchmarks with meshes, where the validity hypothesis behind message-passing is expected to worsen. In these cases, even though the exact message-passing equations do not converge, by introducing a reinforcement parameter to force convergence towards a sub optimal solution, we were able to always outperform the other algorithms with a peak of $27\%$ performance improvement in terms of accommodated paths. On random graphs, we numerically observe two separated regimes: one in which all paths can be accomodated and one in which this is not possible. We also investigate the behaviour of both the number of paths to be accomodated and their minimum total length.
\end{abstract}

\maketitle

\section{Introduction}
The optimization of routing and connection requests is one of the main problems faced in traffic engineering and communication networks \cite{CN}. The need to deliver Quality of service (QoS) \cite{ash,qos2006} performances, when transmitting data over a network subject to overload and failures, requires both efficient traffic management and resource optimization. \\
Some aspects of these problems can be formalized using the edge-disjoint path (EDP) problem. This is a constrained optimization problem that is defined as follows. For a given network and a set of communication requests among pairs of users,  the EDP consists in finding the maximum number of communications that can be accommodated at the same time, under the constraint that different paths cannot overlap on edges. Moreover, the additional requirement of minimization of the total path length can be considered.
Apart from a purely theoretical interest \cite{graphminors}, the EDP finds a wide range of applications: in very-large-scale-integration (VLSI) design, in admission control and virtual circuit routing and in all-optical networks.
 In VLSI design it is required to route wires on a circuit avoiding overlaps, along with minimizing the length of the wires \cite{gerez,chan2003}. In admission control and virtual circuit routing \cite{awerbuch1994,gerla1995,adhoc} one needs to reserve in advance a given path for each communication request so that once the communication is established no interruption will occur. This has applications in real-time database servers, large-scale video servers \cite{shin1994,mahapatra2006,salama1997}, streaming data and bandwidth reservation in communication networks \cite{sumpter,srinivas03,jain2008,Li:2004} and in parallel supercomputers. All these applications require high quality data transmission and full bandwidth exploitation. Routing via edge disjoint paths allows for an efficient bandwidth allocation among users because overlap avoidance means full bandwidth exploitation by each single user.
An area that has attracted particular attention in the last decade is communication transmission in all-optical networks. Along an optical fiber different communications cannot be assigned the same wavelength to transmit data. Moreover, a unique wavelength must be assigned on all the edges contributing to the path assigned for a given communication. Routing communications under the above two requirements define the problem of routing and wavelength assignments (RWA) in this type of networks \cite{RWAreview}. These two constraints suggest that a strategy that iteratively builds edge disjoint paths solutions could allow for a more efficient bandwidth management, namely by using an overall smaller number of wavelengths. This leaves available the remaining ones (according to the edge capacity) to be used either by new users entering the network or by allowing current users to exploit higher bandwidth. This strategy has indeed been applied using greedy \cite{BGA} and genetic algorithms \cite{hsu2014rwa} with performances comparable to other methods based on integer linear programming, graph coloring or bin packing. 
\\
 The EDP is classified among Karp's NP-hard combinatorial problems \cite{karp,garey1979}.  Defining the approximation ratio of a given algorithm as the ratio between the result obtained in term of cost/profit by the algorithm and the optimal one (or viceversa depending on what order gives the maximum ratio),
 the EDP problem is hard to approximate in the worst case; it has been proved that even an approximation with ratio $O(m^{\frac12-\varepsilon})$ is NP-Hard. The best known approximation ratio for the number of accomodated paths is $O(\min\{n^{2/3},\sqrt m\})$ \cite{chekuri2003,erlebach2006}   where $n$ and $m$ are  the number of nodes and edges  in the graph, respectively. 
Negative results on worst-case inapproximability did not stop progress on heuristc approaches. The problem has been studied intensely with a variety of classical techniques:  heuristic greedy algorithms \cite{chen96,BGA,sumpter,srinivas03},  elaborated strategies using bin packing\cite{skorin07}, integer/linear programming relaxations \cite{banerjee1996,kolliopoulos2004,ozdaglar2003routing,baveja2000}, post-optimization \cite{post14}, Montecarlo local search \cite{pham2012}, genetic algorithms \cite{noronha2011biased,gen2008network,DE}, particle swarm optimization \cite{hassan2007dynamic} and ant colony optimization \cite{blesa04}, among them. 

\begin{figure}
\includegraphics[width=4cm]{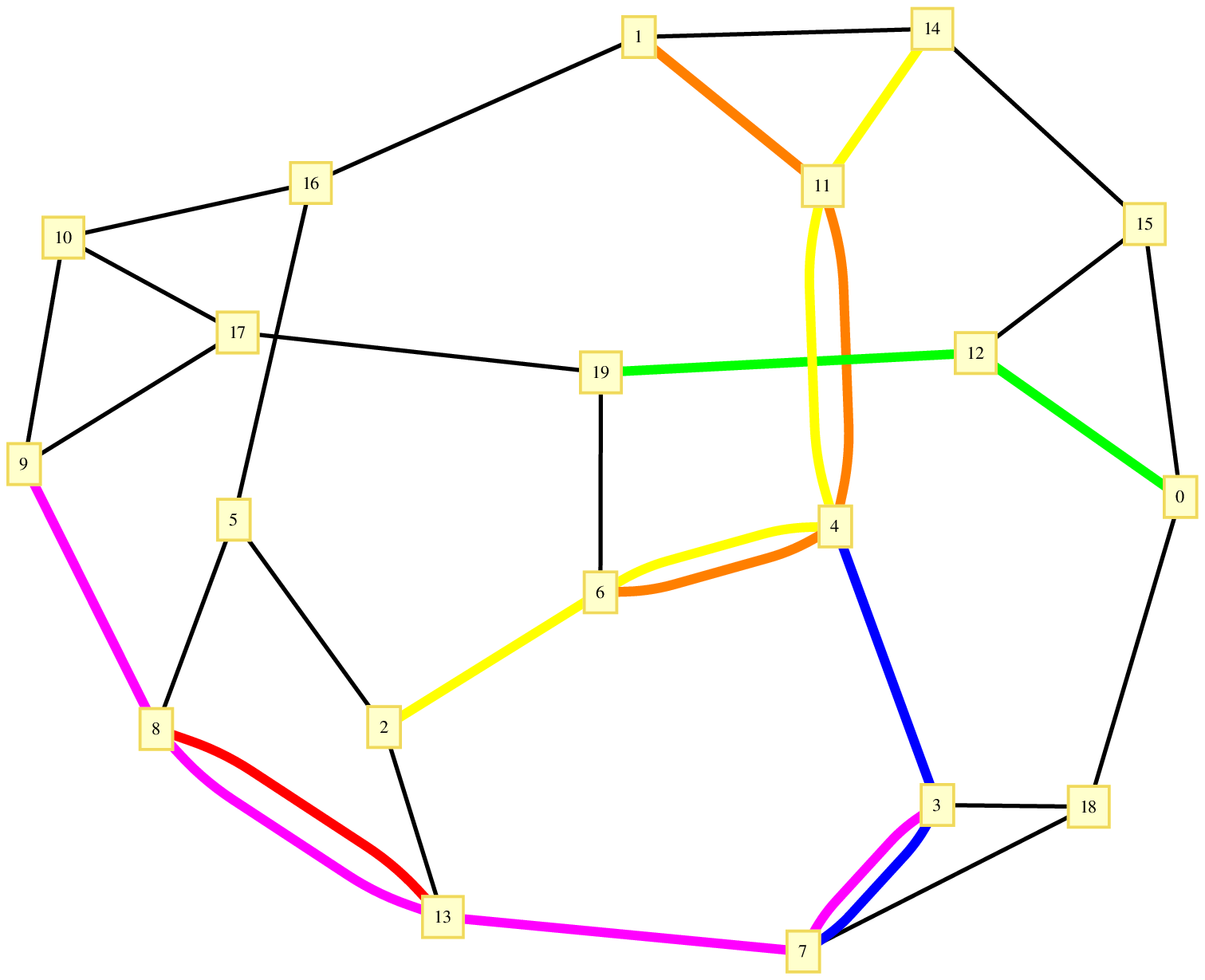}
\includegraphics[width=4cm]{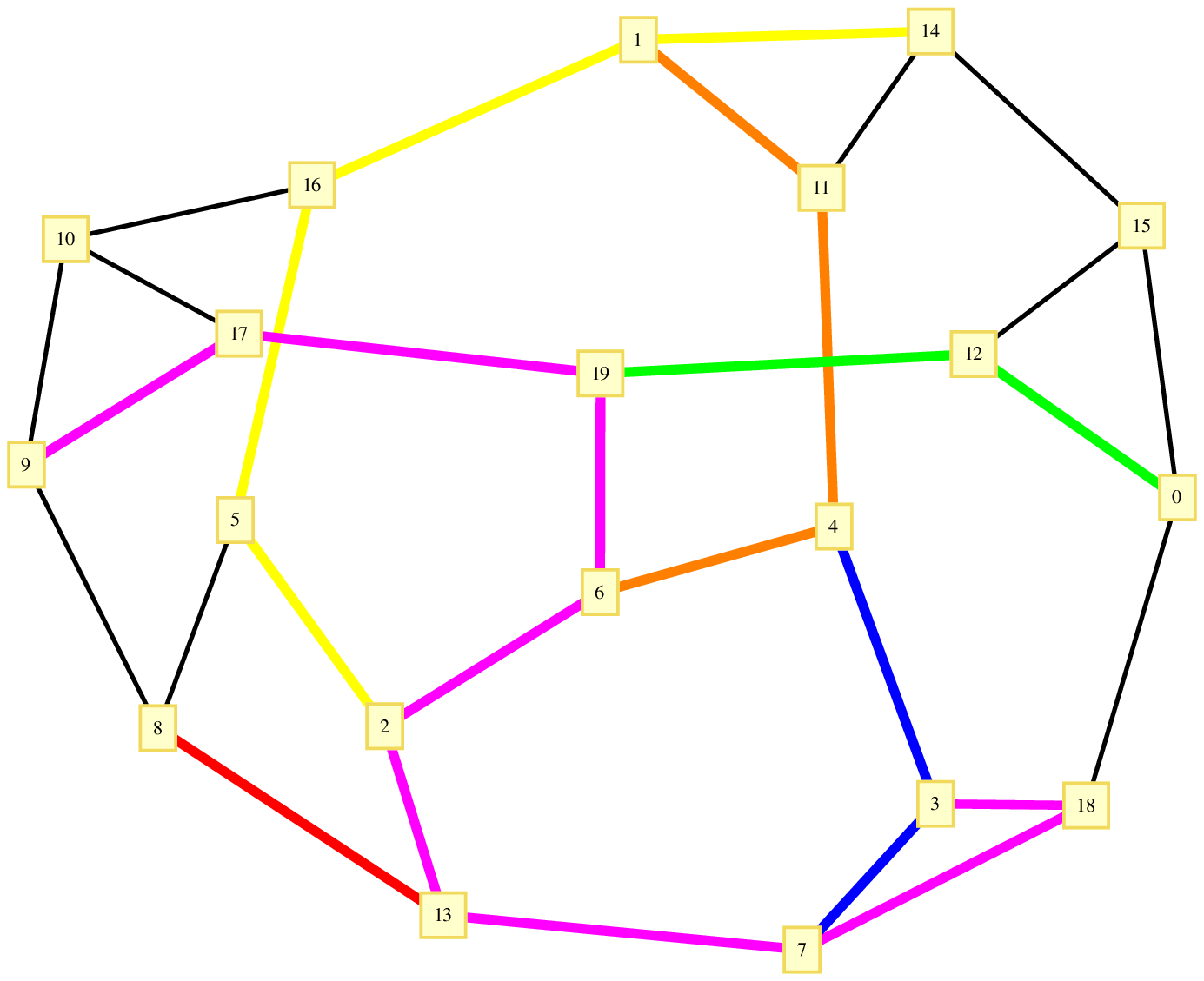}
\caption{An instance of the EDP problem over a 3-regular random graph of $V=20$ and $M=6$: examples of solutions of the unconstrained (left) and optimal (right) EDP problem are displayed. In the latter, the purple communication is redirected along a longer path to avoid edge-overlap. The yellow one has two shortest paths of equal length (degeneracy) in the unconstrained case, but once EDP is enforced this degeneracy is broken and only one of the two is optimal (right).}
\end{figure}


In this paper we propose a distributed algorithm to solve the EDP problem based on message-passing (MP) techniques (or cavity method) \cite{cavity}. This method has been extensively employed to address problems in spin glass theory \cite{sg87,inf09,pd01}, combinatorial optimization \cite{mezard2002} and more recently in routing problems on networks \cite{competition12,pnas13,bayati2008,bacco2014}.
The evaluation of the equations at the core of the MP technique requires, for each vertex $i$ in the underlying graph, to solve a local combinatorial optimization problem, performing a minimum over a set which is exponentially large in the number of neighbors of $i$. We propose an efficient method to perform this calculation, by mapping it into a minimum-weight matching problem on a complete auxiliary graph with vertices in the set $\partial i$ of neighbors of $i$, that can be solved by classical algorithms \cite{matching}.
With this construction, each iteration of the MP equations can be computed in a time which is polynomial in the number of graph edges (and linear in average for sparse random graphs). \\
The MP algorithm is tested on computer-generated instances of different classes of random graphs to study the scaling properties with the system size and to compare the performances against a greedy algorithm.
We also considered the EDP problem on some benchmark instances found in the literature, for which we could compare the message-passing results with those obtained using other types of algorithms: greedy, ant colony optimization \cite{blesa04} and Montecarlo local search \cite{pham2012}.\\ 
The paper is organized as follows. In Section~\ref{sec:Model} we define the EDP optimization problem, for which we present the message-passing equations in Section~\ref{sec:MP}, together with the mapping on a matching problem that simplifies their actual implementation. Section~\ref{sec:rg} reports the results of simulations on random graphs and the scaling of the relevant quantities with the system size, while the comparison between the performances of the message-passing algorithm and other methods is discussed in Section~\ref{sec:comparison}.  Conclusions are given in Section~\ref{sec:conclusion}.

\section{The Edge-Disjoint Paths Problem}
\label{sec:Model}
Given a network and a set of $M$ communications requests between pairs of senders and receivers, the standard EDP consists in finding the maximum number of accommodated paths which are mutually edge disjoint.
In the applications described in the Introduction, the length of the communication paths is a quantity that, directly or indirectly, affects the overall transmission performances, in terms of transmission delays, infrastructure cost and network robustness. We take into account this aspect by considering the Minimum Weight Edge Disjoint Paths (MWEDP) problem, a generalization of the EDP problem that combines in a unique framework both path length optimization and edge disjointness. An instance of the MWEDP problem is defined by a graph $G(\mathcal{V},\mathcal{E})$, where $\mathcal{V}$ denotes the set of nodes and $\mathcal{E}$ is the set of edges, by an assignment of edge weights $w$, that we assume to be non-negative real numbers and by a set of $M$ communication requests $\{(S^{\mu},R^{\mu})\}_{\mu=1,\dots,M}$ between ordered pairs of nodes. We denote by $\pi_\mu$, a path, i.e. a set of consecutive edges $e \in \mathcal{E}$, that connects a sender $S^{\mu}$ with the corresponding receiver $R^{\mu}$. The optimization problem consists in finding $M$ pairwise edge-disjoint paths $\pi_\mu$ while minimizing the total edge weight $\sum_\mu w(\pi_\mu)$, where $w(\pi_\mu)=\sum_{e\in\pi_\mu} w(e)$. 

The classical EDP problem could be trivially recovered by assigning zero weight to all edges in $G(\mathcal{V},\mathcal{E})$ and a positive cost to each communication that is not accommodated. Alternatively, any solution of the MWEDP problem can be reinterpreted as a solution of the classical EDP problem by slightly modifying the original instance of the graph $G(\mathcal{V},\mathcal{E})$ by introducing an extra edge between each pair $(S^{\mu},R^{\mu})$ with sufficiently large cost, such that the algorithm could still always find a solution possibly using these expensive extra edges.  By construction, the cost of each of these $M$ extra edges should be larger than the maximum possible weight a single path can take. Then the solution of the classical EDP problem is obtained from any solution of the MWEDP problem by discarding the paths passing through the extra edges. 
In the present paper, we keep information about path length minimization by assigning unit weights (i.e. $w_{ij}=1, \forall (ij) \in \mathcal{E}$) to the original edges of the graph $G(\mathcal{V},\mathcal{E})$ and a
fixed cost $|\mathcal{E}|+1$ to the extra edge added between each pair $(S^\mu,R^\mu)$.

We introduce $M$-dimensional variables $\bar{I}_{ij}=(I_{ij}^{1}, \dots, I_{ij}^{M})$ with entries $I_{ij}^{\mu} \in \{\pm 1, 0\}$ representing the communication passing along an edge: 
\begin{equation} I_{ij}^{\mu}  = \begin{cases}  1, &  \mbox{if communication $\mu$ passes from $i$ to $j$}, \\
-1, &  \mbox{if communication $\mu$ passes from $j$ to $i$}, \\  0, & \mbox{otherwise}.
\end{cases}
\end{equation}
 We call these vectors \textit{currents} as they must satisfy current conservation at each node $i$ (Kirchhoff law): 
 \begin{equation}
 \sum_{j \in \partial i} I_{ij}^{\mu} + \Lambda_{i}^{\mu}=0, \quad \forall \mu=1,\dots, M,
 \end{equation}
 where we defined for each node $i$ and each communication $\mu$ a variable $\Lambda_{i}^{\mu}$  such that
 \begin{equation}
 \Lambda_{i}^{\mu}= \begin{cases} 1 & \mbox{if $i = S^\mu$,}\\ -1 & \mbox{if $i = R^\mu$,} \\ 0 & \mbox{otherwise}.
 \end{cases}
 \end{equation} 
 The constraint of edge-disjointness specifies that for each edge $(ij)$, at most one of $I_{ij}^\mu$ is non-zero, therefore each vector $\bar{I}_{ij}$ can be parametrized by a variable taking $2M+1$ different values. Notice that the set of variables $\{\bar{I}_{ij}\}_{(ij)\in E}$ completely specifies the state of the network. In this multi-flow formalism, the MWEDP problem is a combinatorial optimization problem in which the global cost function $C(\{\bar{I}_{ij}\})=\sum_{(ij)\in E} w_{ij} f(||\bar{I}_{ij}||)$ depends additively on the total net current $||\bar{I}_{ij}||=\sum_{\mu} |I_{ij}^{\mu}|$ along the edges, and the edge-disjointness is ensured by defining  
\begin{equation}
f(||\bar{I}_{ij}||) = \begin{cases}  0, & \mbox{if $||\bar{I}_{ij}|| =0$}, \\  1, & \mbox{if $||\bar{I}_{ij}|| =1$}, \\ +\infty, & \mbox{if $||\bar{I}_{ij}||  > 1$}, \\ \end{cases}
\end{equation} 
where $|I_{ij}^{\mu}| = 0,1$ denotes the absolute value of $I_{ij}^{\mu}$. Thus configurations with more than one communication passing along an edge have infinite cost and, in the case of unit weights, the total cost $C(\{\bar{I}_{ij}\})$, if finite, represents exactly the total path length, i.e. the number of edges traversed by paths. 

\section{The Message-Passing Algorithm}
\label{sec:MP}
\begin{figure}
\includegraphics[width=5cm]{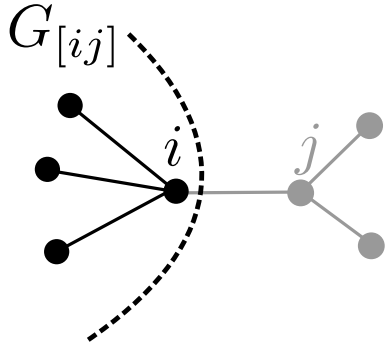}
\caption{The modified cavity graph $G_{[ij]}$.}\label{fig:tree}
\end{figure}

On a tree, the optimization problem defined in Sec \ref{sec:Model} can be solved exactly by iteration using the following message-passing algorithm. Let us assume that $G$ is a tree and consider the subtree $G_{[ij]}$ defined by the connected component of $i$ in $G\setminus (ij)$ (see Figure \ref{fig:tree}). We define $E_{ij}(\bar{I}_{ij})$ to be the minimum cost $C(\{\bar{I}_{ij}\})$ among current configurations that satisfy Kirchoff's laws on all vertices of $G_{[ij]}$ given that we fix an input (or output) extra current $\bar{I}_{ij}$ entering (or exiting) node $i$. Because of the absence of cycles,  it is possible to write a recursive equation for $E_{ij}$ as a sum of cost contributions coming from neighbors of $i$ in the subtree, plus the single cost contribution due to the current $\bar{I}_{ij}$ passing along edge $(ij)$. We call these quantities $E_{ij}(\bar{I}_{ij})$ messages and they verify the \textit{min-sum} recursion relation \cite{inf09}:
\begin{equation}\label{minsum}
E_{ij}(\bar{I}_{ij})= \min_{\{\bar{I}_{ki}\}| constraint}\left\{ \sum_{k \in \partial i \setminus j } E_{ki}(\bar{I}_{ki})\right\} + f(||\bar{I}_{ij}||)
\end{equation}
where \textit{constraint} is the Kirchhoff law at node $i$ and $\partial i$ denotes the neighborhood of $i$. This relation is exact for trees and can be considered as approximately correct for locally tree-like graphs, such as sparse random ones \cite{cavity,inf09}, where correlations between neighbors of a given node decay exponentially. One can develop further this recursion to obtain a set of three types of message-passing equations, one for each type of node, i.e. for each value of $\Lambda_{i}^{\mu}$. 
A fixed point of these equations can be found by iteration from arbitrary initial values for the messages until convergence. Then, one can collect at each edge the incoming and the outgoing converged messages to find the optimal configuration $\{\bar{I}_{ij}^{*}\}_{(ij)\in E}$ such that:
\begin{equation}
\bar{I}_{ij}^{*}= \argmin_{\bar{I}}\left\{E_{ij}(\bar{I}) + E_{ji}(-\bar{I}) - f(||\bar{I}||)\right\}
\end{equation}
where the last term is subtracted to avoid double counting of the cost of the single edge $(ij)$.\\
\subsection{The mapping into a weighted matching problem.}
The min-sum algorithm as in \eqref{minsum} presents a computational bottleneck coming from the fact that for each output current $\bar{I}_{ij}$ there is a large number of possible neighborhood's configurations $\{\bar{I}_{ki}\}_{k\in \partial i \setminus j}$ that are consistent both with the edge-disjoint constraints and with Kirchhoff's law. In the calculation of the minimum in (\ref{minsum}) one needs in fact to consider all possible combinations of paths entering and exiting node $i$; the number of such combinations grows exponentially with the degree of node $i$. Nevertheless, the calculation can be performed efficiently by reducing it to a maximum weight matching problem \cite{matching} on an auxiliary weighted complete graph $G'_{i}$. The nodes of $G'_{i}$ are the neighbors $k \in \partial i$ and the (symmetric) weights matrix $Q$ will be defined as
\begin{equation}\label{weight}
  Q_{kl}=-\min_{1\leq|\nu|\leq M}\left\{E_{ki}(\nu)+E_{li}(-\nu)\right\}+E_{ki}(0)+E_{li}(0)
\end{equation}
where $E_{kl}(\nu)=E_{kl}(\bar{I}_{kl})$ with $I^\mu_{kl}\equiv\delta_{\mu,\nu}$ for $\nu>0$, $I^\mu_{kl}\equiv-\delta_{\mu,\nu}$ for $\nu<0$ and $I^\mu_{kl}\equiv 0$ for $\nu=0$. Notice that this notation maps the $M$-dimensional vectors $\bar{I}_{ij}$ to the $2M+1$ possible current configurations $\nu$ allowed by the edge-disjointness constraint along a given edge. The computation of matrix $Q$, that requires $O(Mk^{2})$ operations, should be performed only once at the beginning of the update routine for node $i \in G$.

\begin{figure}
\includegraphics[width=8cm]{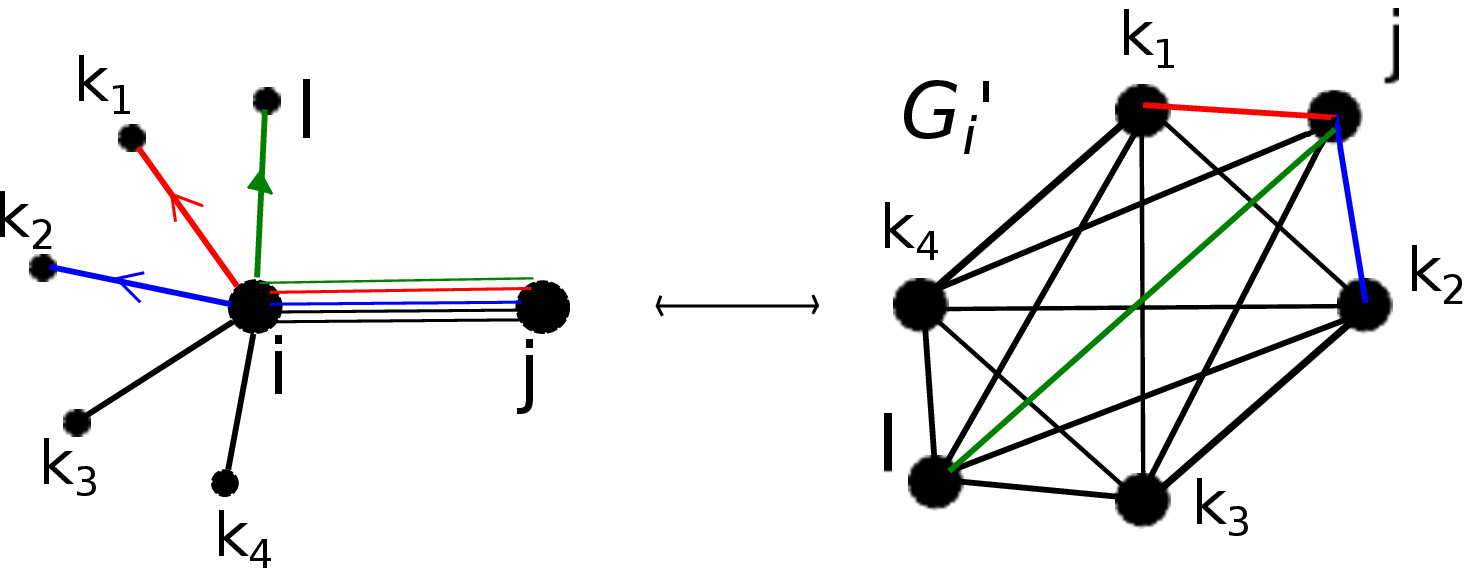} \quad \;
\includegraphics[width=8cm]{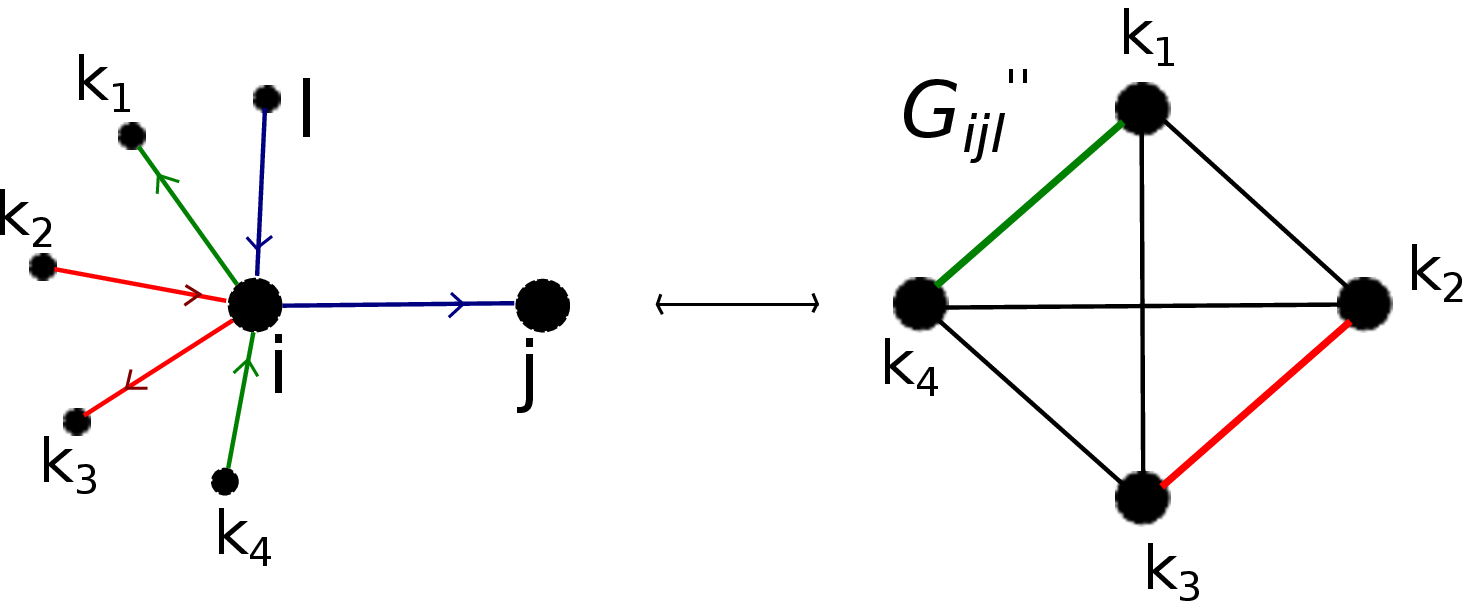}
\caption{Mapping into a weighted matching problem. Left: intermediate step where $G'_{i}$ is built. On the leftmost part we show an example of several communications passing along $(ij)$ and exiting along the remaining neighbors $k \in \partial i \setminus j$. Right: the final step where $G''_{ijl}$ is built; the best configuration around node $i$ when the blue current passes through $(ij)$ is given by the minimum weighted matching on the complete auxiliary graph $G''_{ijl}$. Edges red and green represent the best matching, i.e. the configuration where two other communications enter/exit neighbors of $i \setminus j$.}\label{fig:matching}
\end{figure}

Consider now a neighbor $j \in \partial i$ and a given $\mu$ passing through edge $(ij)$, we want to update $E_{ij}(\bar{I}_{ij})$. Assuming to know the other vertex $l \in \partial i \backslash j$ where the current $\mu$ entering (resp. exiting) node $i$ can exit (resp. enter), then the least costly configuration in the remaining neighborhood is given by 
\begin{equation}
  q^{\min}_{jl} = -M_{jl} + \sum_{k \in \partial i \setminus \{j,l\}} E_{ki}(0)
\end{equation}

where $M_{jl}$ is the maximum weight of a matching on a complete graph ${G''_{ijl}}$ with $k-2$ nodes, built from $G'_i$ by removing nodes $j$ and $l$ (and all their incident edges). Recall that a matching is a subset of edges of $G''_{ijl}$ that do not share any vertex\cite{matching}. This is indeed equivalent to assigning to some of the remaining pairs of neighboring nodes currents $\nu \in [-M, \dots, M]$ that enters through one of them and exits through the other, such that the overall cost of the configuration is minimum. The key point is that the matching condition, i.e. the fact that edges in the solution set cannot have a vertex in common, in our problem translates in the condition of forbidding edge overlaps. Hence, thanks to this auxiliary mapping, we are able to reduce the computation of the update rule for the MP equations of the edge disjoint path problem to the solution of a standard (polynomial) combinatorial optimization problem, i.e. maximum weight matching. In Figure \ref{fig:matching} we give a diagrammatic representation of the mapping. Note that in the maximum matching problem, edges in the input graph with negative weight can be simply removed.
  Notice that the neighboring current $\nu$ can also be a priori equal to $\mu$ in this algorithm, because the configurations where $\mu$ appears in more than one pair of edges will be eliminated in the minimization calculation as they have higher cost in our formulation. The minimum weight is thus independent of $\mu$, i.e. of which message we are updating, a fact that allows reducing the complexity of the algorithm by a factor $M$. 
  
  Finally one needs to minimize over $l$ given the matrix $q^{\min}$:
\begin{equation}\label{eq:matching}
E_{ij}^{t+1}(\mu)= \min_{l \, \in \, \partial i \, \backslash j} \left \{E_{li}^{t}(\mu) + q^{\min}_{jl} \right \} + c_{ij}(\mu) 
\end{equation}
where $c_{ij}(\mu)$ is the cost of edge $(ij)$, that in our case is $0$ if $\mu=0$ and $1$ otherwise.
We can notice that in order to evaluate each term inside the brackets we need to perform a matching optimization on each of the $(k-2)$-node complete graphs $G''_{ijl}$ built $\forall \,l \, \in \, \partial i \, \backslash j$. Each of these matching routine has complexity $O(k^{3}\log k) $ \cite{galil1986} and there are $O(k^{2})$ possible combinations of $j$ and $l$. 
Reminding that we first need to evaluate the weight matrix $Q$, the overall complexity of this algorithm will be:
$$ O(k^{5} \log k + Mk^{2})$$
which is polynomial in the variables $k$ and $M$. Once we have performed this whole procedure, we get all the information we need to calculate the $2M+1$ update messages $E_{ij}^{t+1}(\mu)$, for each $j \, \in \, \partial \, i$, adding a term $O(kM)$ to the final complexity (which is nonetheless negligible compared to the previous two).

The case of $\mu=0$, in which no current passes through edge $(ij)$ regardless of what happens on the other edges, is addressed by calculating a matching on the $(k-1)$-node complete graph composed of all nodes $l \, \in \partial \, i \backslash j$.
If $i$ is either a sender or a receiver, i.e. $\Lambda_{i}^{\mu}\in\{ \pm 1\}$ for a given $\mu \in [1,\dots, M]$, the same computation can be performed provided that an auxiliary node, indexed by the communication label $\mu$ is added to the original graph $G$ and connected to node $i$, such that its exiting messages will be fixed once at the beginning in the following way and never updated: $E_{\mu i}(\nu)=-\infty$ if $0<\nu=\mu$ (sender) or $0<-\nu=\mu$ (receiver), and $E_{\mu i}(\nu)=+ \infty$ otherwise.

\subsection{The role of reinforcement.}
\begin{figure}[htbp]
\centering
\includegraphics[width=0.6\columnwidth]{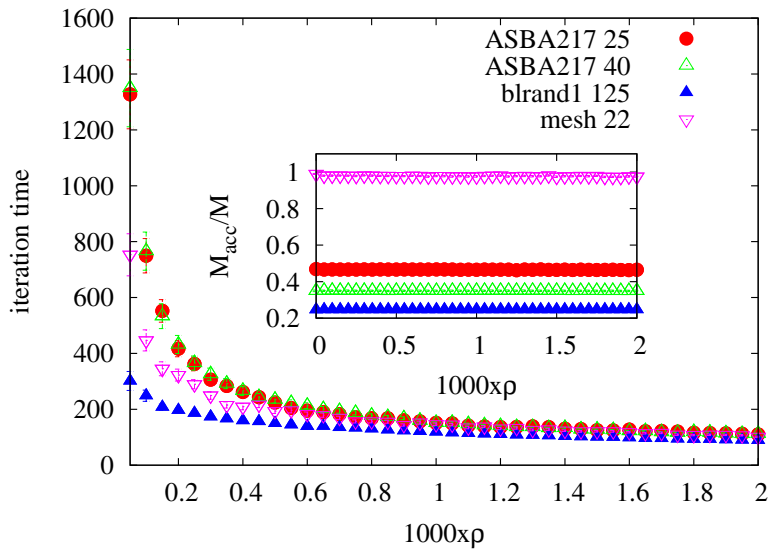}
\caption{Reinforcement performance. Number of iterations to reach convergence as a function of reinforcement parameter $\rho$ on BRITE graphs AS-BA217 (V=100) with $M=25,40$, blrand1 (V=500) with $M=125$ and mesh 15x15 with $M=22$. Inset: the number of accomodated paths $M_{acc}$ is substantially unchanged in the range of parameter values under study.}\label{fig:rein}
\end{figure}

\begin{figure}[!htbp]
\centering
\includegraphics[width=0.49\columnwidth]{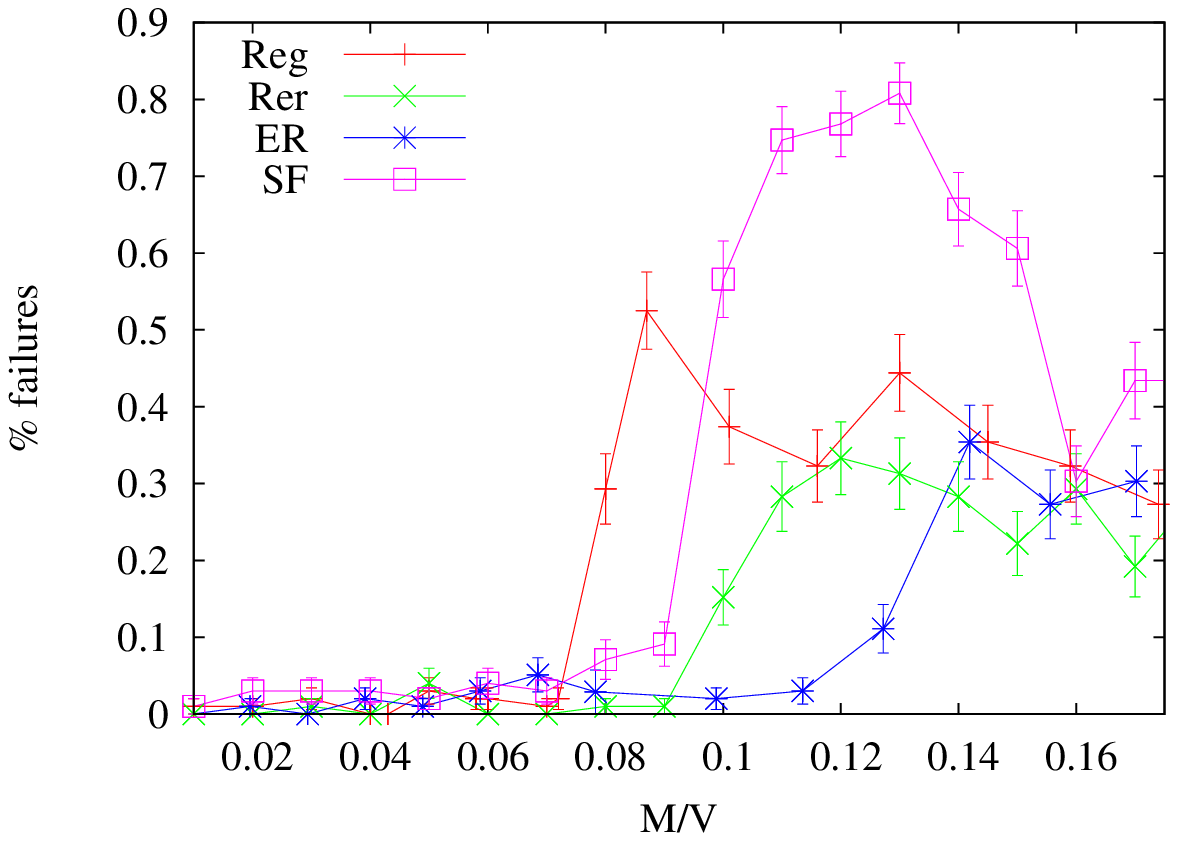} 
\includegraphics[width=0.49\columnwidth]{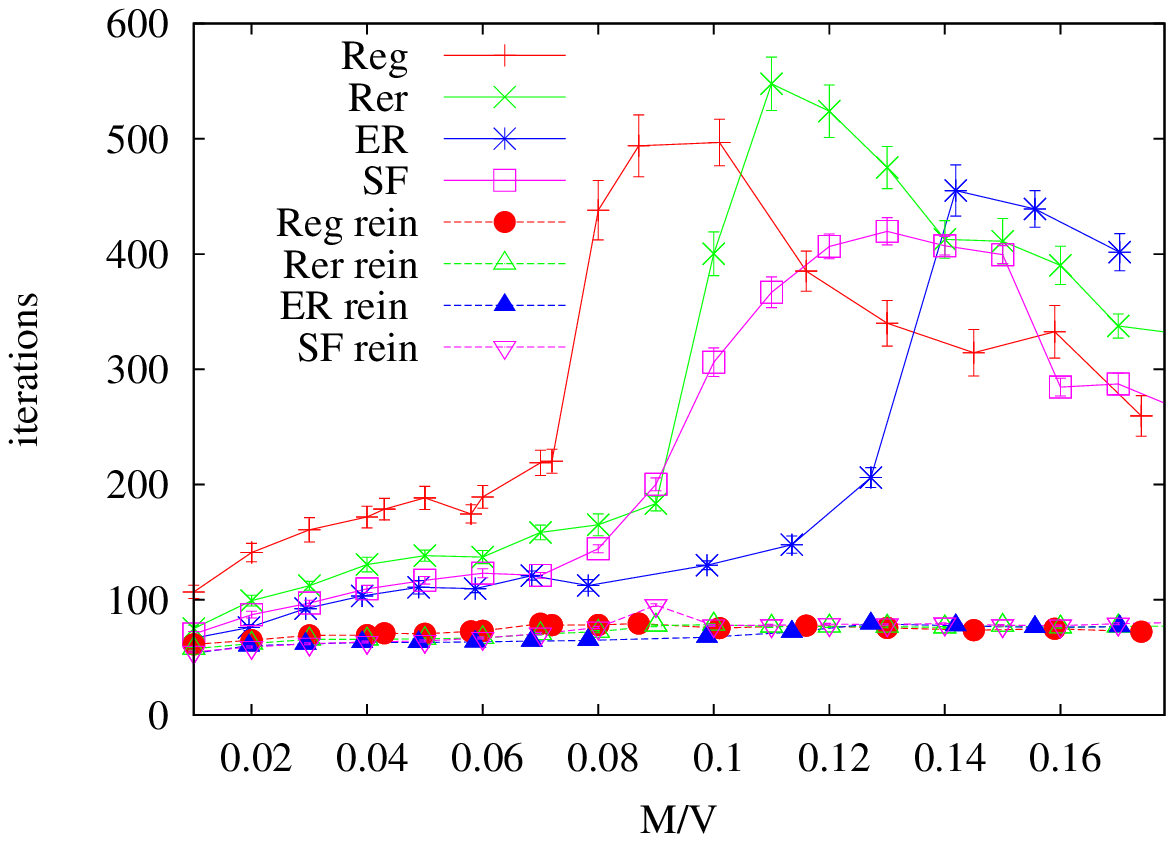}
\caption{Left: Fraction of instances in which convergence standard MP fails (reinforced MP always converged in our experiments). Right: number of iterations for convergence for standard MP and reinforced MP ($\rho=0.002$) in case of random graphs of $V=1000$ and $\langle k \rangle =3$ as a function of $M/V$. Notice how the reinforcement term, besides ensuring convergence, greatly improves the convergence time.
}
\label{convergence}
\end{figure}

In order to aid and speed-up convergence of the MP equations, we used a reinforcement technique \cite{braunstein2006, altarelli2009}, in which a set of external local fields $h_{ij}^{t}(\mu)=E_{ij}^{t}(\mu)+E_{ji}^{t}(-\mu)-c_{ij}(\mu)$ act on the messages gradually biasing them to align with themselves. The reinforcement is introduced by promoting edge costs to become communication-dependent quantities defined as linear combinations of the cost at the previous time-step and the reinforcement local fields:
\begin{equation}
c^{t+1}_{ij} (\mu)= c^{t}_{ij}(\mu)+\gamma_t h_{ij}^{t}(\mu) 
\end{equation}
with $c_{ij}^{0}(\mu)=c_{ij}$. This cost will then be inserted into equation (\ref{eq:matching}) to replace the  term $c_{ij}(\mu)$. This has the effect to lead the messages to converge faster, gradually bootstrapping the system into a simpler one with large external fields. In practice we choose $\gamma_t = t \rho$ and one has to choose the growth rate of $\gamma$ by tuning the reinforcement parameter $\rho$, that controls the trade-off between having a faster convergence and reaching a better solution. We tested $\rho$ on instances on three types of graphs to finally choose to fix it to $\rho=0.002$ in the rest of the simulations. In Figure \ref{fig:rein} we could notice that this value achieves comparable results (inset) in terms of $M_{acc}/M$ to lower $\rho$ in less time. \\
In Figure \ref{convergence}(left) we report the number of converged instances (over 100 realizations) for standard MP (without reinforcement) on four types of random graphs (as described in the next section) and fixed size $V=1000$ and average degree $\langle k \rangle =3$. The convergence failure of the standard MP increases considerably with $M/V$ until it reaches a peak value, then it decreases. On the contrary, when reinforcement is used, convergence is always achieved in less than $100$ steps (right panel). 

\section{Results on random graphs}\label{sec:rg}
\begin{figure}[!htbp]
        \centering
        \begin{tabular}{@{}cc@{}}
          \includegraphics[width=0.5\columnwidth]{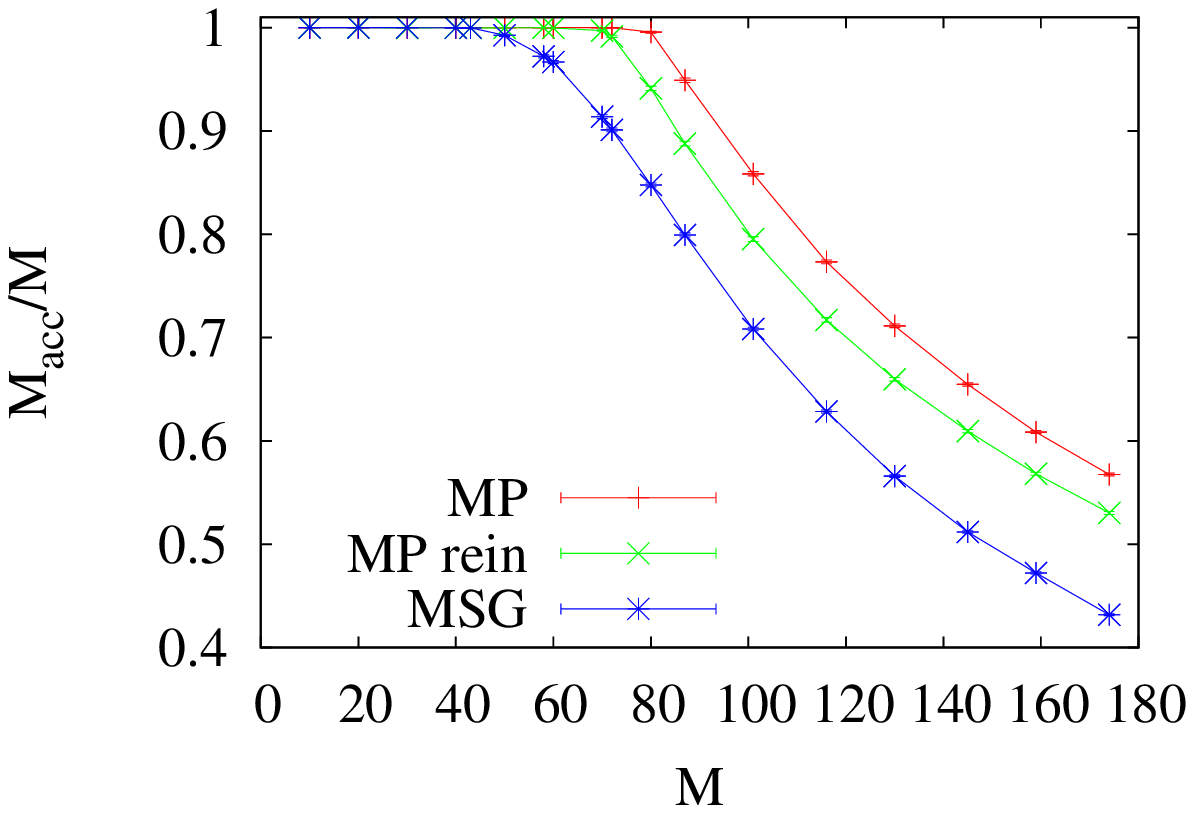} & \includegraphics[width=0.5\columnwidth]{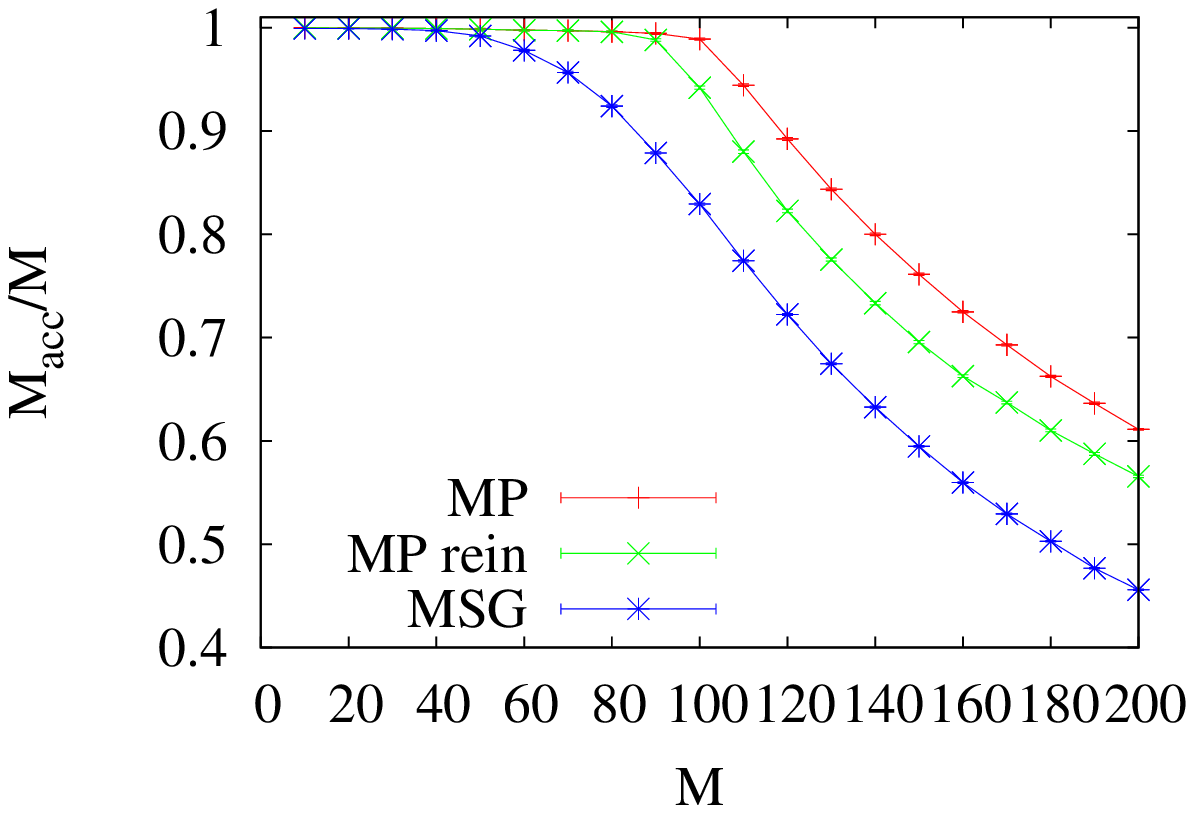}\\
          \includegraphics[width=0.5\columnwidth]{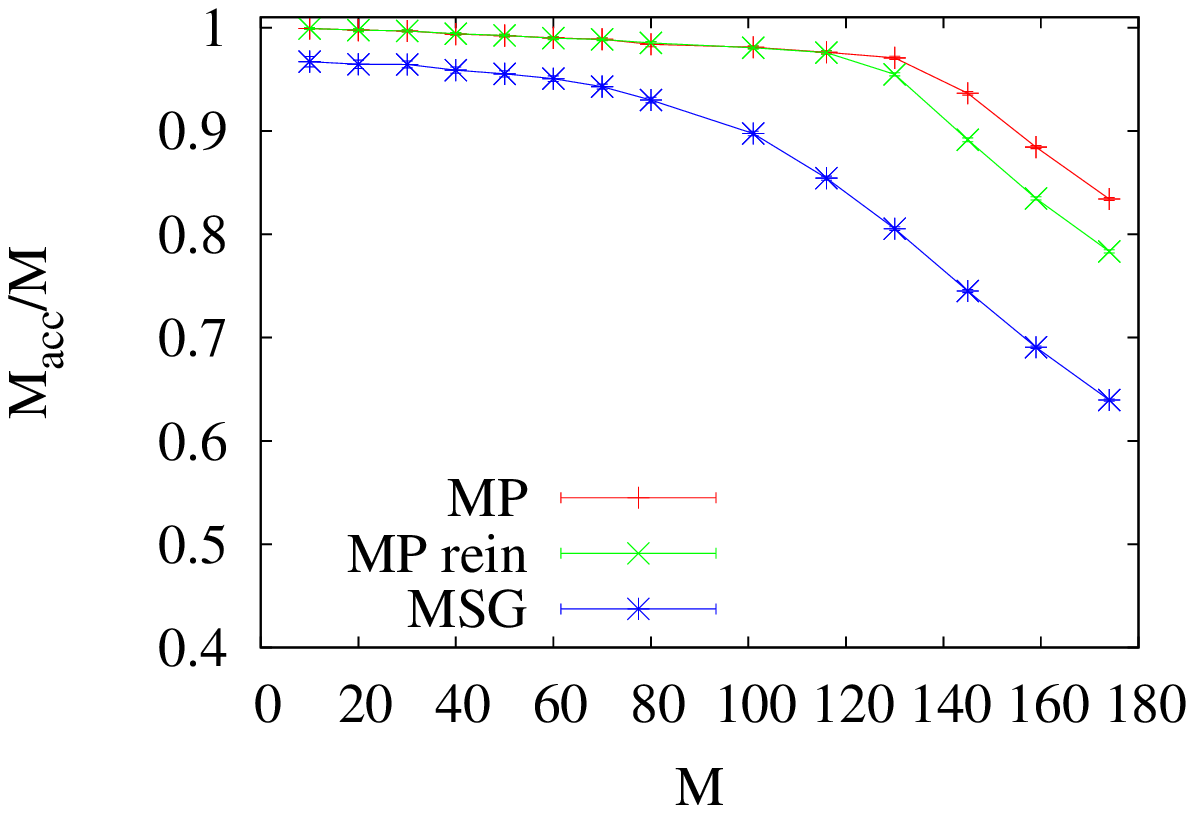} & \includegraphics[width=0.5\columnwidth]{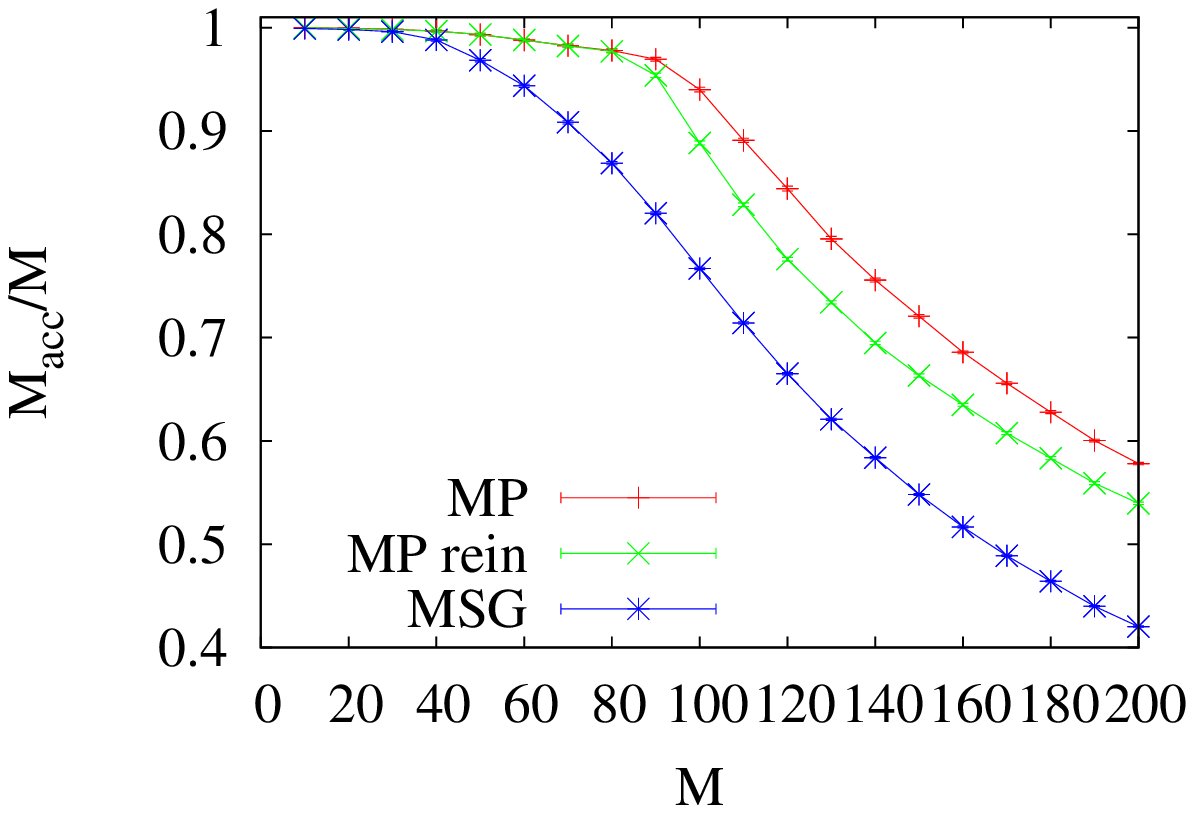}
           \end{tabular}
\caption{MP vs greedy performance. We plot the performance in terms of $M_{acc}/M$ for (from top to bottom) regular, RER, ER and SF graphs of fixed size $V=10^{3}$ and average degree $\langle k \rangle=3$. Error bars are  smaller than the size of the symbols. 
}\label{fig:Maccomodated}
\end{figure}

\begin{figure}[!htbp]
        \centering
        \begin{tabular}{@{}cc@{}}
          		\includegraphics[width=0.4\columnwidth]{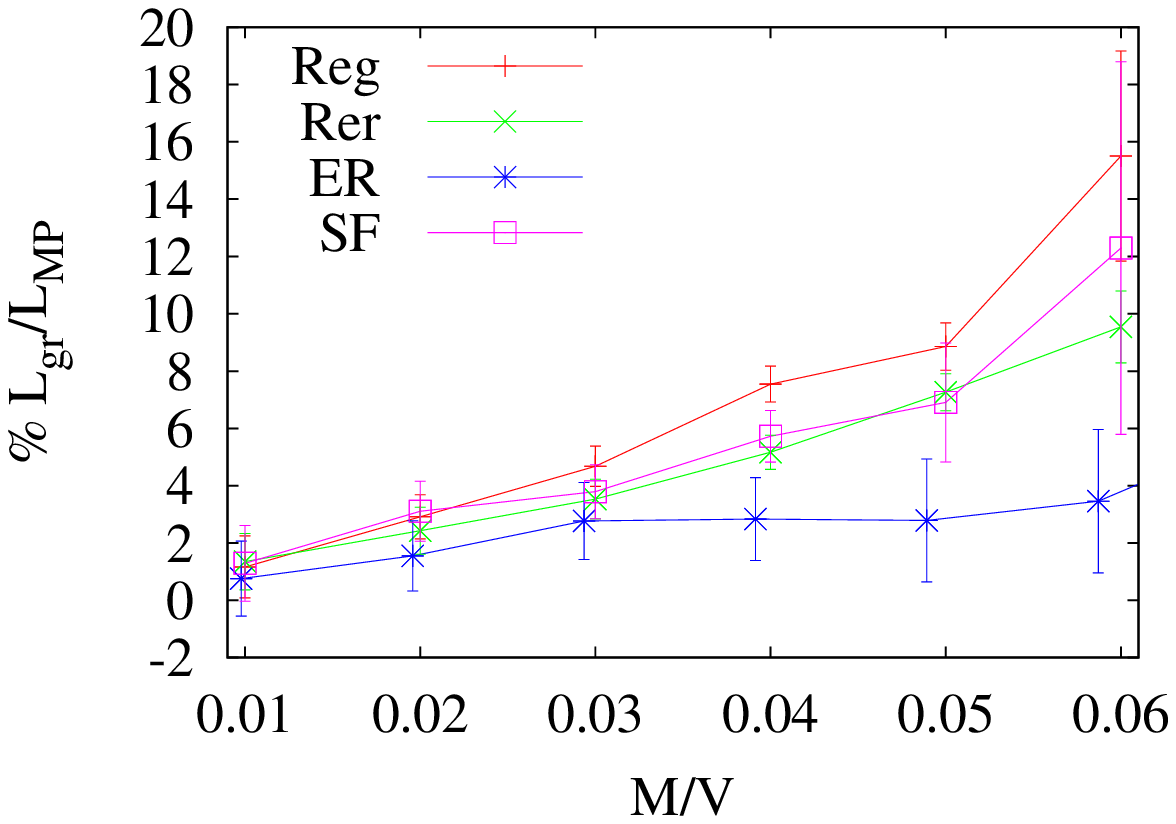} & \includegraphics[width=0.4\columnwidth]{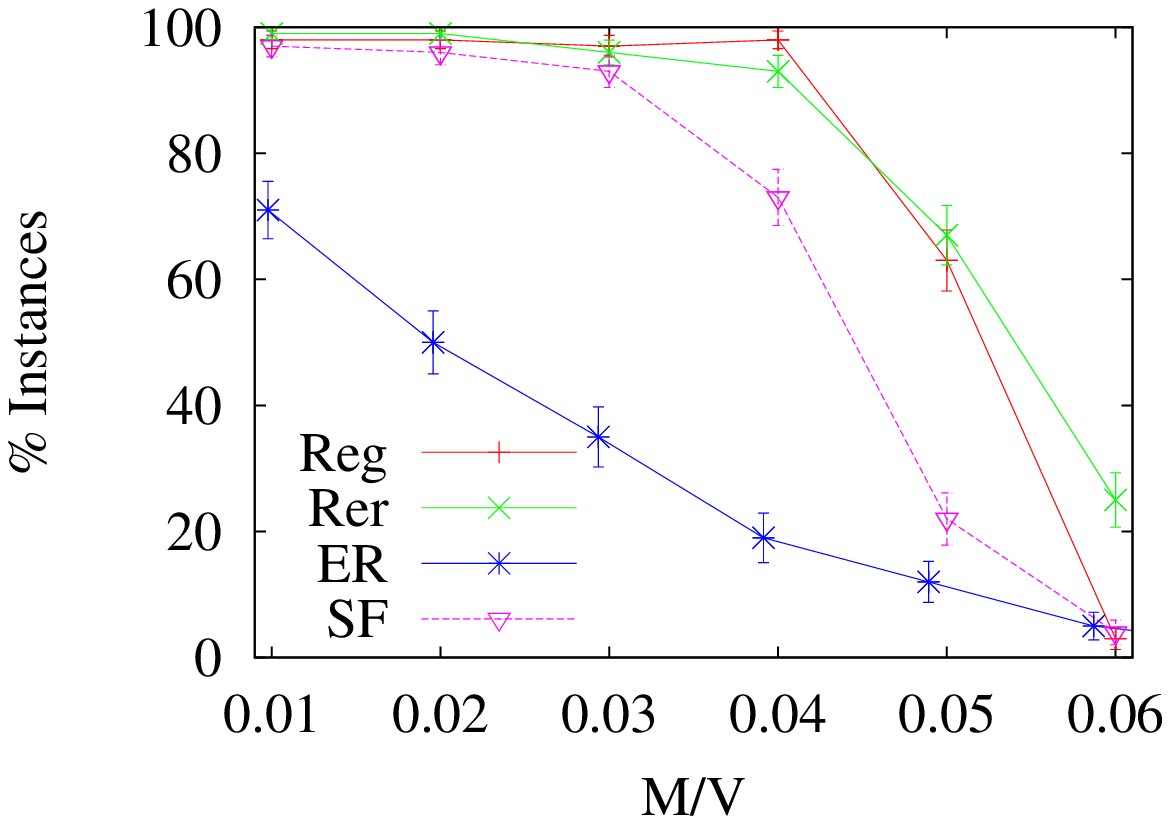} \\
               		\includegraphics[width=0.4\columnwidth]{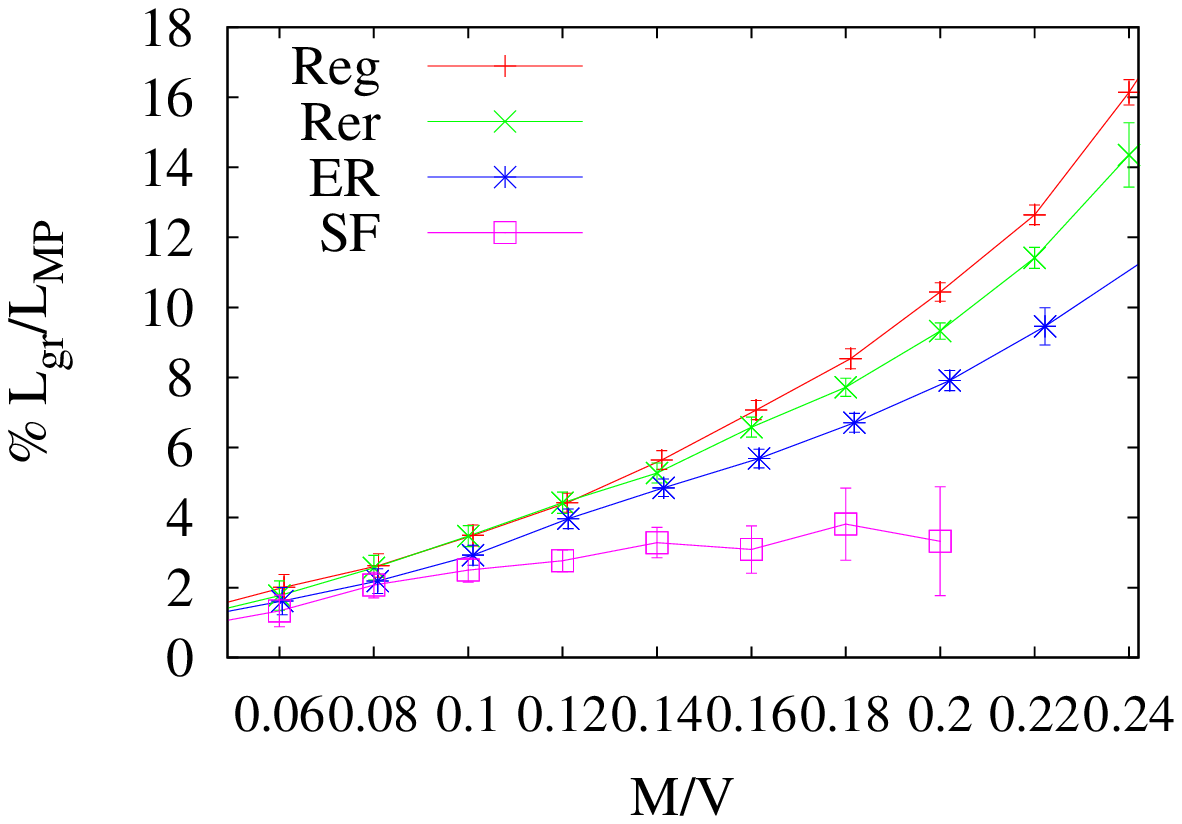} & \includegraphics[width=0.4\columnwidth]{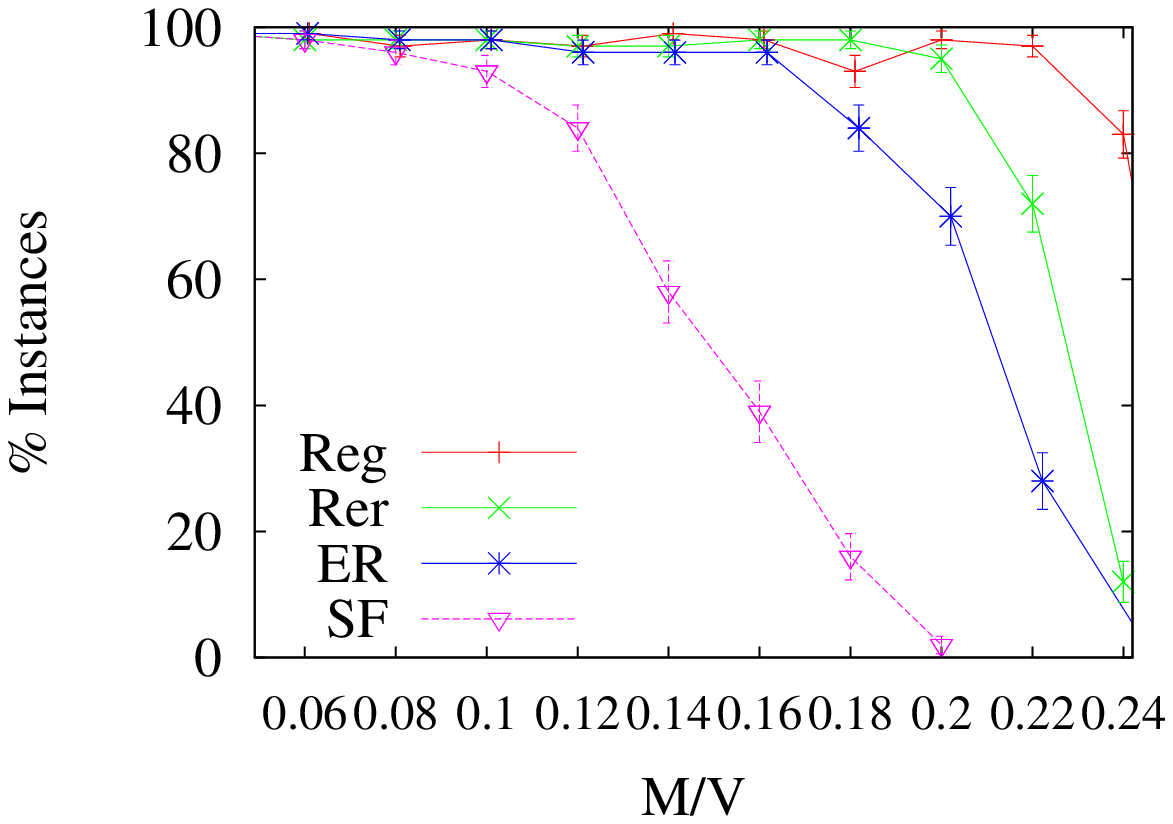} \\
               		\includegraphics[width=0.4\columnwidth]{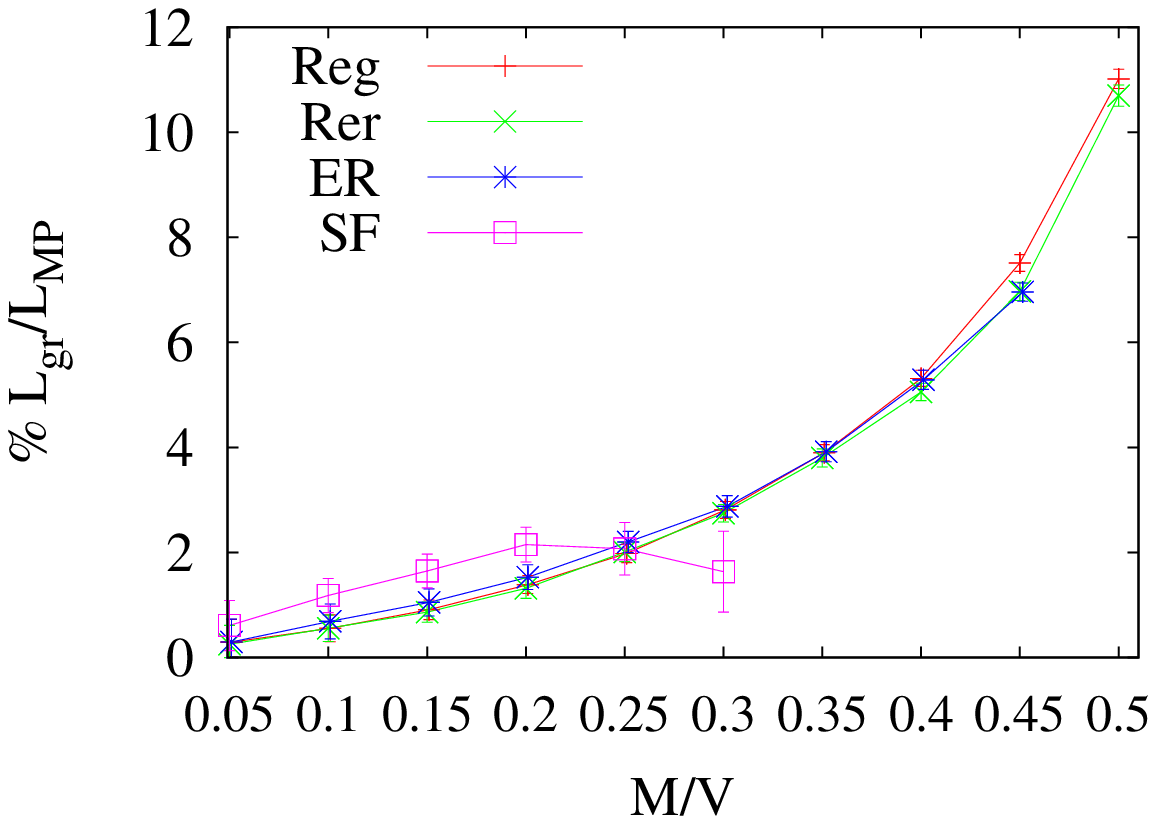} & \includegraphics[width=0.4\columnwidth]{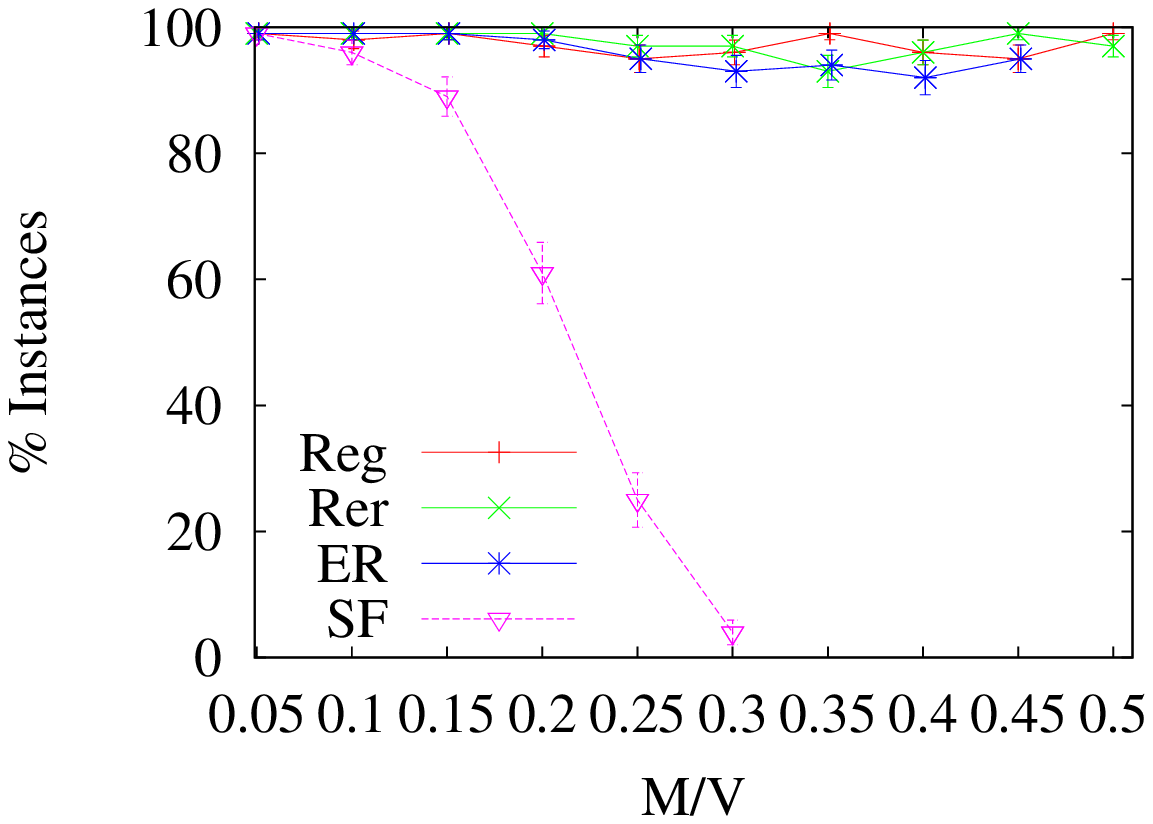}
           \end{tabular}
\caption{Length performance. We plot (left) the relative performance of MSG over MP in terms of total length of the solution paths: $y=100(L_{g}/L_{MP}-1)$. Here $L_{g}$ and $L_{MP}$ denote the total path lengths calculated with MSG and MP respectively. We use Reg, RER, ER and SF graphs of fixed size $V=10^{3}$ and average degree $\langle k \rangle=3,5,7$ (from top to bottom). On the right we report the number of instances where the two algorithms find the same solution in term of $M_{acc}/M$ over 100 realizations.}
\label{fig:comparison}
\end{figure}

\begin{figure}[!htbp]
        \centering
          \begin{tabular}{@{}cc@{}}
          \includegraphics[width=8cm]{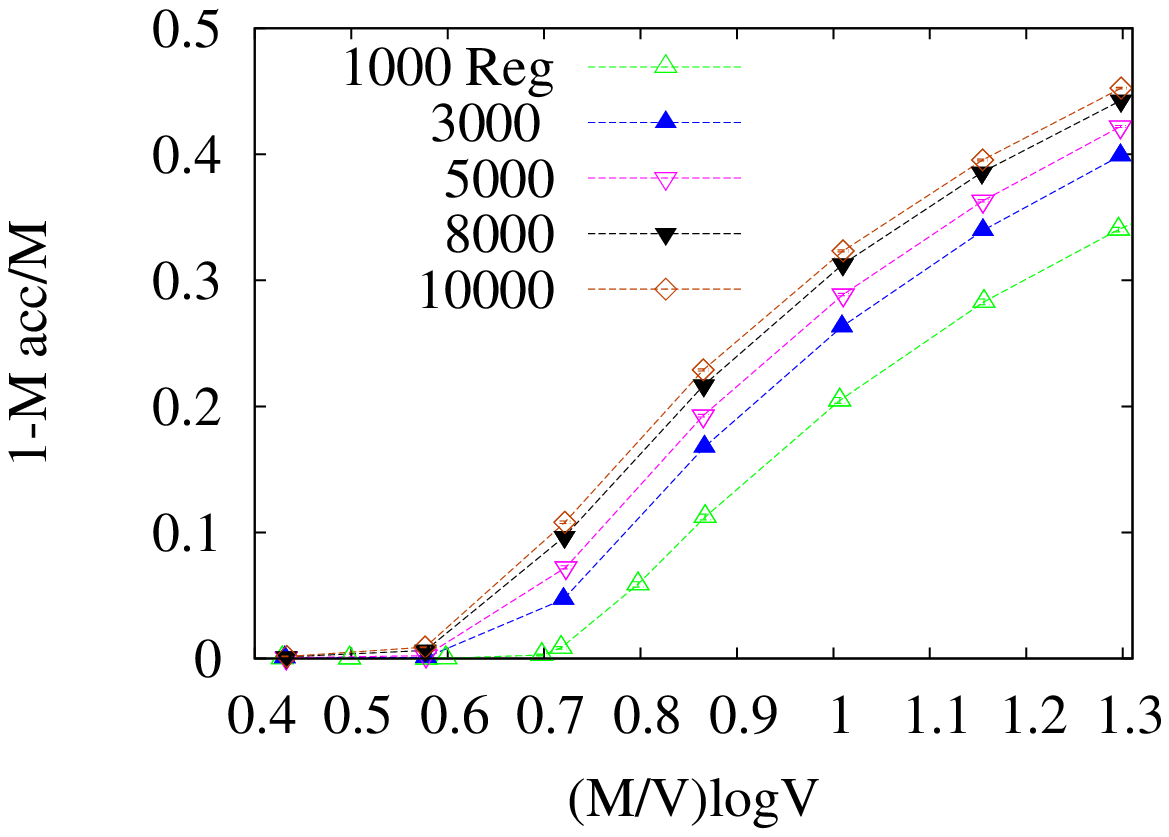} &
                    \includegraphics[width=8cm]{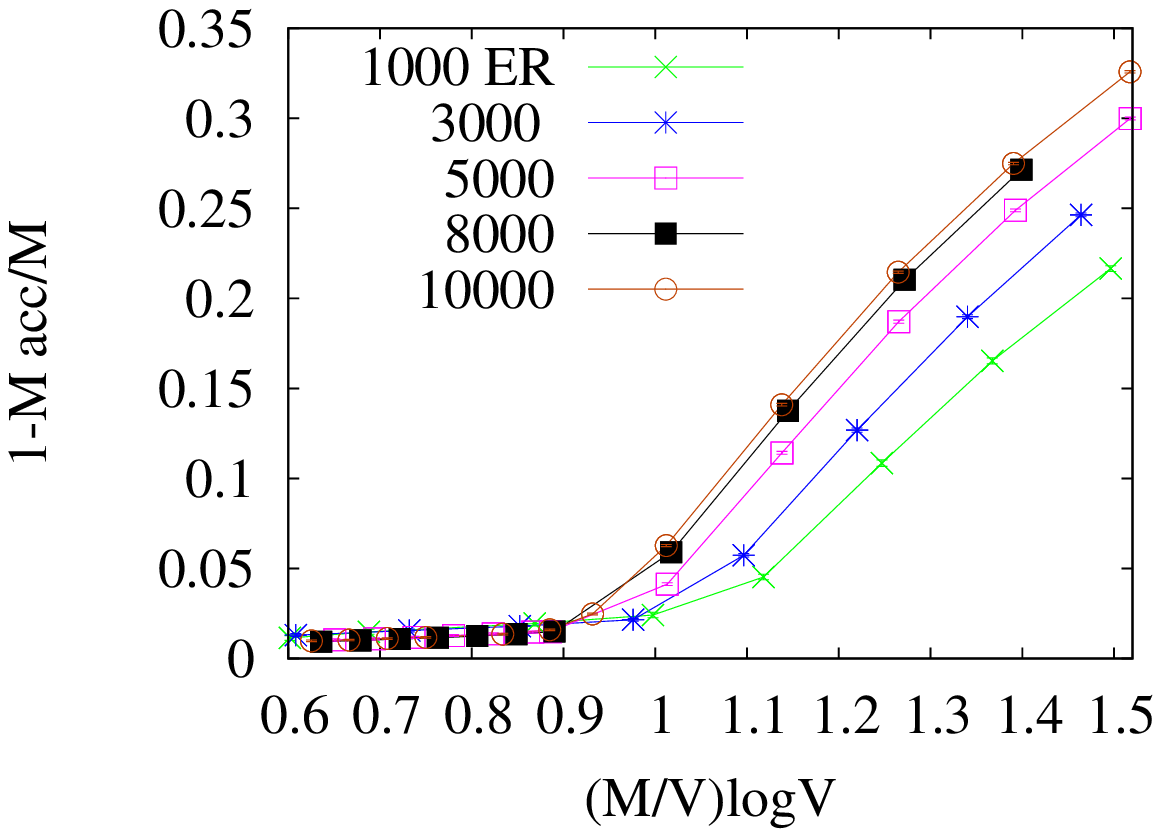} \\
                      \includegraphics[width=8cm]{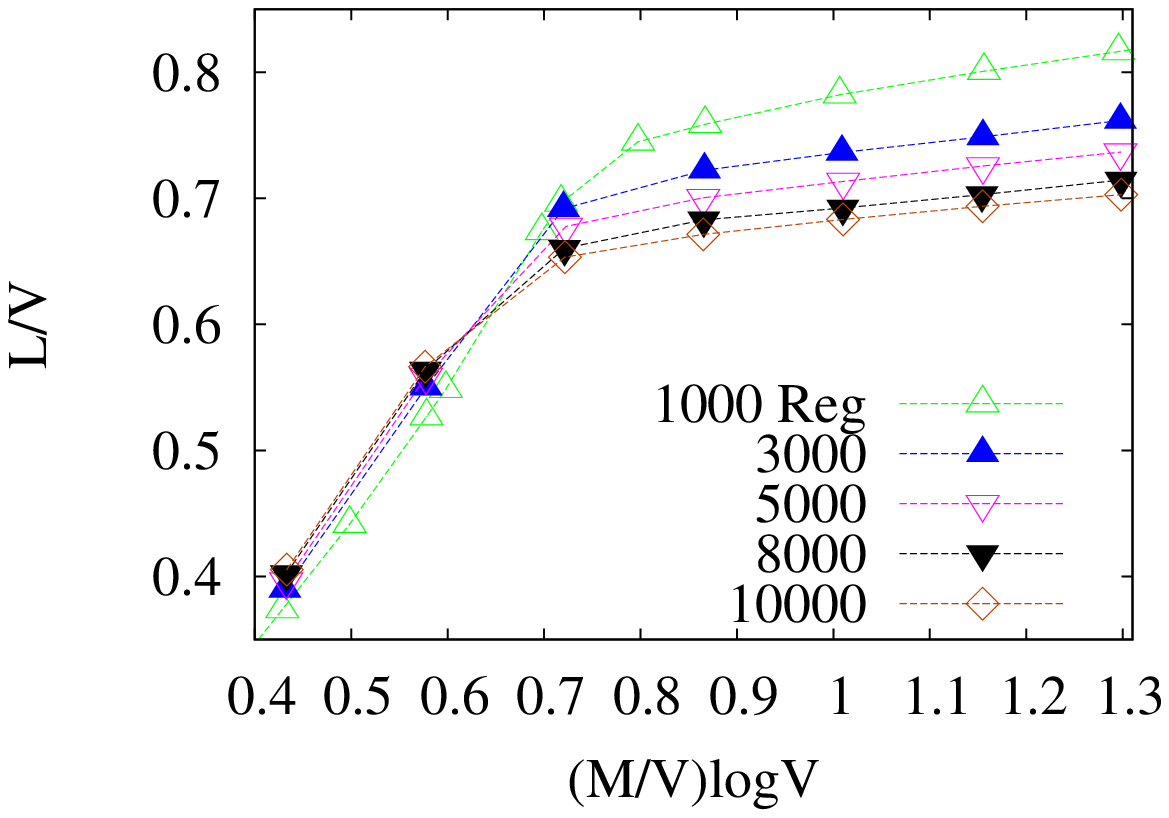} &
                     \includegraphics[width=8cm]{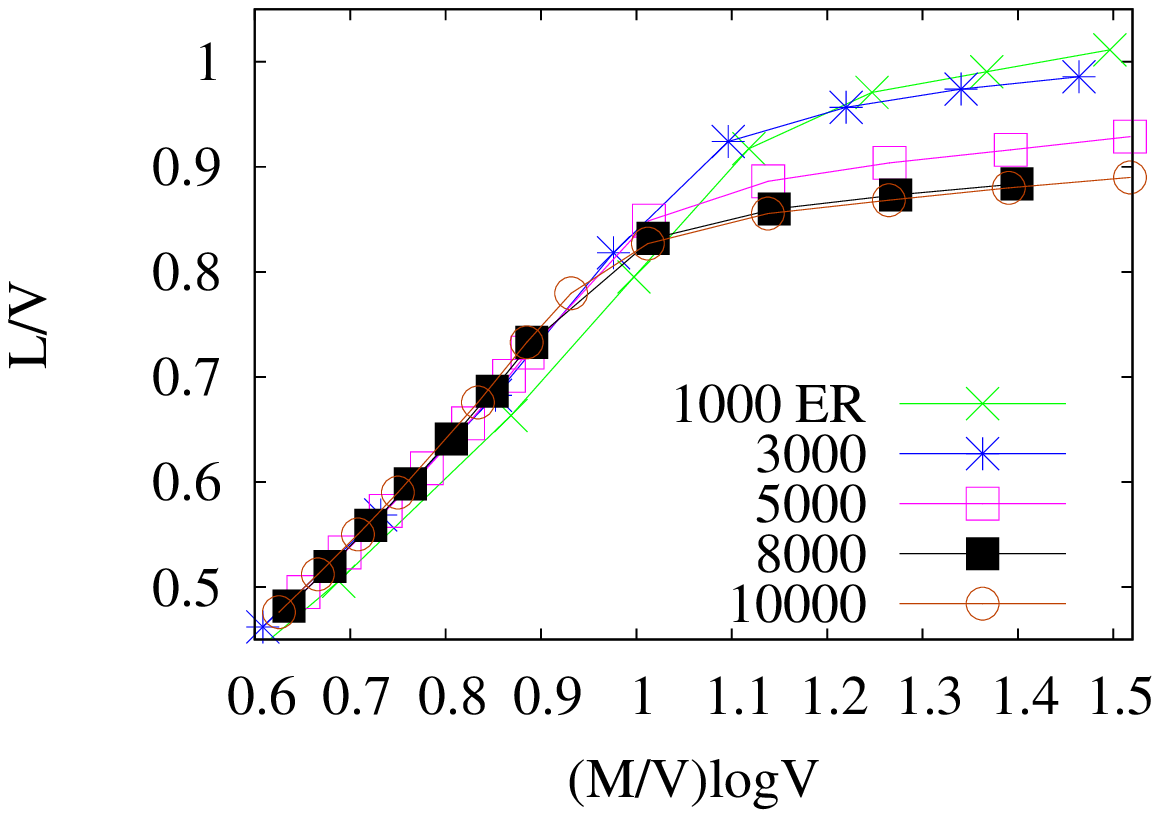}
                             \end{tabular} 
\caption{Finite-size effects. We plot $1- M_{acc}/M$ (top) and the total length per node $L/V$ (bottom) for Reg (left) snd ER (right) graphs as a function of the scaling variable $\frac{M\log V}{V}$. We can notice the finite-size effects decreasing with system size leading to the curves corresponding to the biggest graphs $V=8000,\, 10000$ to almost superimpose. Note that in the SAT phase the total length grows linearly in $\log V$ for all system sizes as expected but in the UNSAT phase the graphs split. Error bars are smaller than point size.}
\label{scaling}
\end{figure}

First, we tested the MP algorithm on various types of random graphs, with fixed size $V= |\mathcal{V}| =1000$ and average degree $\langle k \rangle=3, 5, 7$: regular random graphs (Reg), Erd\H{o}s-R\'{e}nyi random graphs (ER) \cite{ER}, random graphs with power-law distribution (SF) \cite{SF} and a set of graphs (RER) obtained adding edges independently with probability $p$ starting from a $k_{0}$-regular random graph (for large $V$, the final average degree of such graphs is $\langle k \rangle =k_{0} + d$, with $d=pV$). We compared the performance with a multi-start greedy algorithm (MSG) \cite{blesa04}. This heuristic algorithm calculates paths by iteratively choosing a (random) communication $\mu$, finding the corresponding shortest path and removing the edges belonging to the path from the graph. The process is repeated until either there are no paths left to be routed or no communications can be accommodated anymore in the graph. The multi-start version repeats the same procedure a given number of times and keeps the best solution in terms of $M_{acc}$, the number of accommodated paths. A bounded-length version \cite{kleinberg} of MSG has been used to develop an iterative algorithm to solve the RWA using EDP in \cite{BGA}: its performance was comparable to the one obtained using a linear programming solver on graphs of small sizes ($V\leq 40$) but with faster execution times.
This makes it suited to be tested on larger graphs. A disadvantage of the greedy method is that it relies heavily on the order in which communications are accommodated (it disregards the information about sender-receiver pairs other than the ones already accommodated). The difference in the performances of the message-passing and greedy algorithm could then be used to assess the relevance of local information usage in such optimization problem. We tested both the standard multi-start and the bounded-length version but we found equal results with the first being slightly faster, in our tests, in terms of execution times. Thus we decided to use the standard MSG in our simulations.
First we compared the results in terms of number of accommodated paths $M_{acc}$ by calculating the ratio $M_{acc}/M$. In Figure \ref{fig:Maccomodated} we show the behavior of $M_{acc}/M$ for each type of random graph and $V=10^{3}$, $\langle k \rangle =3$ using MP, reinforced MP and MSG. Both MP versions perform better than MSG, with the standard MP giving better results. The corresponding results for $\langle k \rangle=5$ are similar (not reported) but the value $M_{acc}/M<1$ is reached at higher values of $M/V$ and standard MP and MP with reinforcement give almost always the same solutions. The case $\langle k \rangle=7$ is not reported because, given the high number of edges, the solutions are often trivial (i.e. $M_{acc}/M=1$), a part from the case of SF graphs where we have instead $M_{acc}/M<1$ due to the presence of many small degree nodes. We also studied the total path length as a function of $M/V$ for the solutions, obtained with the different algorithms. We consider the ratio between the total path lengths obtained with greedy and MP for solutions in which the number $M_{acc}$ of accommodated path is the same. In Figure \ref{fig:comparison} we can see that MP always outperforms the MSG algorithm for all types of graph under study. The results for the SF graph with $\langle k \rangle =7$ are quite different from the other graphs: both for MP and MSG the ratio departs from $1$ at rather small values of $M/V$, possibly because the maximum number of accommodated paths is limited by the existence of many small degree nodes that act as bottlenecks, preventing the use of many alternative edge-disjoint routes.
The scaling behavior of the fraction $1-M_{acc}/M$ of unaccommodated communications and the average total path length $L/V$ of accommodated paths with the system size in the solutions obtained using the MP algorithm  is shown in Figure \ref{scaling} for regular random graphs and ER random graphs. These quantities are plot as functions of the scaling variable $x=\frac{M\log V}{V}$. Note that when paths do not interact, $x$ is a measure the total path length per site, as the average path lenght is proportional to $\log V$. In the top panels, two regimes are visible: for small $x$, all communications can be accommodated, whereas  at some value $x^*$ the curves for different values of $V$ depart from zero. This behavior can be interpreted as a SAT/UNSAT transition, in analogy with the terminology of constraint-satisfaction problems \cite{mezard2002}. The collapse of the curves $L/V$ for different values of $V$ is very good in the region in which all paths can be accommodated. On the contrary, in the UNSAT region, the curves for different sizes do not collapse anymore, though the relative difference between them seems to decrease by increasing the system size, and the curves for the largest graphs analyzed ($V=8000,\,10000$) are almost superimposed. We argue that $x$ is the correct scaling variable in the limit of infinitely large graphs, and the observed mismatch could be due to finite-size effects. The change of slope in the roughly linear behavior of the average total length $L/V$ is motivated by the fact that in the SAT region, all communications can be accommodated at the cost of taking longer paths with respect to those actually accommodated in the UNSAT region. 


\section{Comparison with other methods} \label{sec:comparison}
A comparison between the performances of the MP algorithm and those of alternative algorithms proposed in the literature \cite{blesa04,pham2012}  is reported in Table \ref{table:benchmark}. As benchmark instances we used: two internet-like topologies generated using the BRITE graph generator \cite{brite} with parameters set as in \cite{blesa04}; mesh graphs of sizes 15x15 and 25x25, Steiner and planar graphs as reported in \cite{pham2012}. For each of these graphs we used the same set of sender-receiver pairs of size $M=0.10V, 0.25V, 0.40V$ used in \cite{pham2012}. For each of these instances we ran the MP, MP with reinforcement and MSG algorithms $20$ times and collected the average, minimum and maximum number of accommodated paths $M_{acc}$ along with the average computational time in seconds. All results are reported in Table \ref{table:benchmark}.

\subsection{Other optimization methods.}
A part from the multi-start greedy, we used as comparison two more structured algorithms. The first one is an Ant Colony Optimization metaheuristic \cite{blesa04}. This method builds an EDP solution incrementally from partial solutions provided by a set of $M$ ants. Each ant generates a path for a given communication making probabilistic decisions during the construction steps. These are made by processing local information modeled as \textit{pheromone} information provided by other ants. The advantage of this method is to divide the EDP in subproblems and to use local information. The drawback is that it relies on several parameters that need to be carefully tuned in order to have a sensitive solution. Moreover the computational time increases considerably with the system size. 
The second algorithm is a Montecarlo-based Local Search \cite{pham2012}, that uses as main Montecarlo step a path rewiring based on rooted spanning trees. Unfortunately the running time grows rapidly with the system size, making it computationally expensive when used on large graphs. Results are reported in Table \ref{table:benchmark}.
Finally, we performed simulation using the multi-start greedy heuristic described above. 

\subsection{Results.}
In Table \ref{table:benchmark} we report the performance comparison in terms of $M_{acc}$ between the two versions of MP (with and without reinforcement) and the other 3 types of algorithms. The message-passing always performs equal or better than the other methods. Surprisingly the best performances are given for meshes and planar graphs, where we would expect the failure of MP due to the existence of short loops. What we find instead is that, even though the standard MP converges in few of these instances on meshes, the version with reinforcement  always finds a solution that is always better than the other algorithms.  The larger performance gap is seen on larger set of commodities and bigger graphs. Performance improvement reaches $27\%$ with respect to LS, the best one between the other algorithms tested. The same considerations can be made in the case of planar graphs. We claim that this gap would increase with system size, but unfortunately the size of benchmark graphs remains limited to $V\leq 500$. Moreover these alternative algorithms do not consider path length optimization, thus we cannot compare the performance with respect to this variable. The ACO has been recently tested on several types of graphs (still with $V \leq 500$) against a Genetic Algorithm (GA) in \cite{hsu2014}. It performed better than GA in the case of BRITE graphs 1-6 and  $14\%$ worse in the case of 10x10 and 15x15 mesh graphs. The MP algorithm always outperforms ACO and in the case of 15x15 mesh the gap reaches $23.5\%$. Unfortunately neither the GA has been tested on larger graphs nor gives results in terms of path length of the solutions. 

\section{Conclusions}
\label{sec:conclusion}
The EDP problem is a combinatorial optimization problem that finds applications in several traffic engineering problems, from VLSI design to routing and access control management in communication networks. In this work we proposed a min-sum message-passing algorithm to find the maximum number of communications $M_{acc}$ that can be accommodated in a network subject to edge-disjoint constraints and minimizing total path length at the same time. We devised an efficient method to implement these equations by exploiting a mapping into a minimum weight matching problem on an auxiliary graph.
The standard MP algorithm and the version with reinforcement consistently outperform alternative algorithms found in the literature on different types of benchmark graphs in terms of the fraction $M_{acc}/M$ of accommodated communications. We found two different behaviors: on some ``easy" instances, all algorithm accommodate all requests, providing the same results and suggesting that these could be the optimal ones;  there are non-trivial instances in which $M_{acc}/M<1$, but the message-passing algorithm always outperforms the other algorithms in terms of the number of accommodated paths. In particular we obtained better results in the case of meshes and planar graphs, even though these topologies are not locally tree-like as required by the cavity method. In these cases, we could always ensure convergence of the MP equations by exploiting a reinforcement technique. The quality of solutions improves with decreasing the reinforcement parameter, such that we could always find better solutions than those obtained using the other algorithms under study.
Unfortunately, for the heuristic algorithms employed on the benchmarks we could not access other relevant metrics such as the average total path length, as it was not considered before in the literature \cite{blesa04,pham2012}. Nonetheless we could directly compute such quantity for a multi-start greedy heuristic in several graphs, finding that MP always gives a lower average path length for solutions with the same fraction of accommodated communications.

In conclusion, combining the good performance results, in terms of traffic and path length,  with the polynomial time implementation, the use of the MP algorithm opens new perspectives in the solution of relevant routing problems over communication networks such as the RWA in optical networks. In particular, it would be interesting to apply the MP algorithm in the iterative construction of RWA solutions over communication networks with finite link capacity, as it has been done for other types of EDP algorithms.

\section*{Acknowledgement}
We thanks P. Quang Dung for kindly providing us the detailed benchmark instances.
This work was supported by the Marie Curie Training Network NETADIS (FP7, grant 290038). L.D. also acknowledges the Italian FIRB Project No. RBFR10QUW4.

\section*{References}
\bibliography{bibliography}{}
\bibliographystyle{unsrt}

\appendix

\section{Convergence criterion}
 Given a decision variable $d^{t}$ to be calculated at each iteration update step $t$, an integer variable $n$ and a time step $T_{max}$ we have convergence if, for $n$ consecutive iteration steps, $d^{t}$ does not change, and we fix a maximum iteration time $T_{max}$ to update MP equations. Formally this writes:
\begin{equation}
\exists \, t_{0} \in [1, T_{max}-n] \quad s. t. \quad d^{t_{0}+i}=d^{t_{0}} \quad \forall i=1,\dots, n
\end{equation}
In our simulations we defined the decision variable as the total difference of the optimal currents (calculated edge by edge) between two consecutive iteration steps :
\begin{equation}
d^{t}=\sum_{(ij)\in E}[ 1-\delta_{\mu_{ij}^{t},\mu_{ij}^{t-1}}]
\end{equation} 
where $\mu_{ij}^{t}= |\min_{\mu=-M,\dots, M}\left\{E_{ij}(\mu)+E_{ji}(-\mu)-c_{ij}(\mu) \right\}|$ and convergence is reached when $d^{t}=0$ for $n$ consecutive time steps.

\section{Benchmark results}

\begin{table*}[tp]
\resizebox{0.9\linewidth}{!}{%
\centering
\begin{tabular}{rrrr|rrr|rrr|rrr|rrr|rrr|rrr}
\multicolumn{4}{c|}{Instance}  & \multicolumn{ 3}{c|}{MP} & \multicolumn{ 3}{c|}{MP rein = 0.002} & \multicolumn{ 3}{c|}{MSG (greedy)} & \multicolumn{ 3}{c|}{ACO} & \multicolumn{ 3}{c|}{LS} &  \multicolumn{3}{c}{MP gain vs.} \\ 
\multicolumn{1}{c}{Name} & \multicolumn{1}{c}{$|V|$} & \multicolumn{1}{c}{$|E|$} & \multicolumn{1}{c|}{$\langle k \rangle$} &
\multicolumn{1}{c}{$\langle M \rangle$} & \multicolumn{1}{c}{$M_{min}$} & \multicolumn{1}{c|}{$M_{max}$} &
\multicolumn{1}{c}{$\langle M \rangle$} & \multicolumn{1}{c}{$M_{min}$} & \multicolumn{1}{c|}{$M_{max}$} &
\multicolumn{1}{c}{$\langle M \rangle$} & \multicolumn{1}{c}{$M_{min}$} & \multicolumn{1}{c|}{$M_{max}$} &
\multicolumn{1}{c}{$\langle M \rangle$} & \multicolumn{1}{c}{$M_{min}$} & \multicolumn{1}{c|}{$M_{max}$} &
\multicolumn{1}{c}{$\langle M \rangle$} & \multicolumn{1}{c}{$M_{min}$} & \multicolumn{1}{c|}{$M_{max}$} &
\multicolumn{1}{c}{MSG} & \multicolumn{1}{c}{ACO} & \multicolumn{1}{c}{LS} \\ \hline
blrand10\_1 &500 &1020 &4.08 &16.00 &16 &\bf 16 &16.00 &16 &\bf 16 &13.65 &13 &15 &14.80 &14 &\bf 16 &16.00 &16 &\bf 16 &6.67 & 0.00 & 0.00 \\ \hline
blrand25\_1 &500 &1020 &4.08 &32.00 &32 &\bf 32 &32.00 &32 &\bf 32 &27.75 &26 &30 &31.85 &31 &\bf 32 &32.00 &32 &\bf 32 &6.67 & 0.00 & 0.00 \\ \hline
blrand40\_1 &500 &1020 &4.08 &38.00 &38 &\bf 38 &38.00 &38 &\bf 38 &33.10 &32 &35 &37.85 &37 &\bf 38 &37.90 &37 &\bf 38 &8.57 & 0.00 & 0.00 \\ \hline
blrand10\_2 &500 &1020 &4.08 &26.00 &26 &\bf 26 &25.65 &25 &\bf 26 &23.85 &23 &25 &25.25 &25 &\bf 26 &26.00 &26 &\bf 26 &4.00 & 0.00 & 0.00 \\ \hline
blrand25\_2 &500 &1020 &4.08 &35.00 &35 &\bf 35 &35.00 &35 &\bf 35 &30.75 &29 &33 &34.75 &34 &\bf 35 &34.95 &34 &\bf 35 &6.06 & 0.00 & 0.00 \\ \hline
blrand40\_2 &500 &1020 &4.08 &37.00 &37 &\bf 37 &37.00 &37 &\bf 37 &32.45 &31 &34 &36.95 &36 &\bf 37 &36.95 &36 &\bf 37 &8.82 & 0.00 & 0.00 \\ \hline
blsdeg10\_1 &500 &1020 &4.08 &17.00 &17 &\bf 17 &16.89 &15 &\bf 17 &14.65 &14 &16 &15.95 &15 &16 &17.00 &17 &\bf 17 &6.25 & 6.25 & 0.00 \\ \hline
blsdeg25\_1 &500 &1020 &4.08 &36.00 &36 &\bf 36 &36.00 &36 &\bf 36 &31.55 &30 &33 &35.80 &35 &\bf 36 &36.00 &36 &\bf 36 &9.09 & 0.00 & 0.00 \\ \hline
blsdeg40\_1 &500 &1020 &4.08 &34.00 &34 &\bf 34 &34.00 &34 &\bf 34 &29.00 &28 &31 &33.65 &33 &\bf 34 &34.00 &34 &\bf 34 &9.68 & 0.00 & 0.00 \\ \hline
blsdeg10\_2 &500 &1020 &4.08 &20.00 &20 &\bf 20 &19.85 &19 &\bf 20 &16.90 &16 &18 &19.20 &19 &\bf 20 &20.00 &20 &\bf 20 &11.11 & 0.00 & 0.00 \\ \hline
blsdeg25\_2 &500 &1020 &4.08 &34.00 &34 &\bf 34 &34.00 &34 &\bf 34 &28.45 &27 &30 &32.95 &32 &\bf 34 &33.90 &33 &\bf 34 &13.33 & 0.00 & 0.00 \\ \hline
blsdeg40\_2 &500 &1020 &4.08 &37.00 &37 &\bf 37 &37.00 &37 &\bf 37 &31.75 &30 &33 &36.50 &35 &\bf 37 &37.00 &37 &\bf 37 &12.12 & 0.00 & 0.00 \\ \hline
mesh15\_10\_1 &225 &420 &3.73 &22.00 &22 &\bf 22 &22.00 &22 &\bf 22 &20.60 &20 &\bf 22 &19.65 &19 &21 &21.55 &21 &\bf 22 &0.00 & 4.76 & 0.00 \\ \hline
mesh15\_25\_1 &225 &420 &3.73 &36.00 &36 &\bf 36 &35.10 &35 &\bf 36 &28.30 &27 &30 &27.70 &26 &29 &32.00 &31 &33 &20.00 & 24.14 & 9.09 \\ \hline
mesh15\_40\_1 &225 &420 &3.73 &43.00 &43 &\bf 43 &42.50 &42 &\bf 43 &30.10 &28 &32 &35.30 &32 &38 &38.80 &37 &40 &34.38 & 13.16 & 7.50 \\ \hline
mesh15\_10\_2 &225 &420 &3.73 &- &- &- &19.89 &19 &\bf 20 &19.75 &19 &\bf 20 &17.50 &17 &19 &19.45 &19 &\bf 20 &0.00 & 5.26 & 0.00 \\ \hline
mesh15\_25\_2 &225 &420 &3.73 &35.00 &35 &\bf 35 &34.70 &33 &\bf 35 &29.25 &29 &30 &29.20 &28 &31 &33.05 &32 &34 &16.67 & 12.90 & 2.94 \\ \hline
mesh15\_40\_2 &225 &420 &3.73 &42.00 &42 &\bf 42 &41.35 &41 &\bf 42 &29.80 &29 &32 &34.00 &33 &36 &37.60 &36 &39 &31.25 & 16.67 & 7.69 \\ \hline
mesh25\_10\_1 &625 &1200 &3.84 &- &- &- &47.25 &46 &\bf 48 &40.70 &40 &42 &32.85 &29 &36 &41.00 &39 &43 &14.29 & 33.33 & 11.63 \\ \hline
mesh25\_25\_1 &625 &1200 &3.84 &- &- &- &68.30 &67 &\bf 69 &48.40 &47 &51 &45.00 &42 &49 &55.55 &54 &59 &35.29 & 40.82 & 16.95 \\ \hline
mesh25\_40\_1 &625 &1200 &3.84 &- &- &- &88.74 &88 &\bf 90 &54.35 &53 &58 &57.70 &53 &61 &69.30 &67 &72 &55.17 & 47.54 & 25.00 \\ \hline
mesh25\_10\_2 &625 &1200 &3.84 &- &- &- &44.33 &43 &\bf 46 &40.05 &38 &42 &30.10 &28 &33 &37.90 &36 &40 &9.52 & 39.39 & 15.00 \\ \hline
mesh25\_25\_2 &625 &1200 &3.84 &- &- &- &67.22 &65 &\bf 70 &48.90 &47 &52 &45.60 &44 &48 &54.70 &52 &59 &34.62 & 45.83 & 18.64 \\ \hline
mesh25\_40\_2 &625 &1200 &3.84 &- &- &- &88.55 &87 &\bf 90 &54.05 &51 &57 &57.75 &54 &61 &68.85 &66 &71 &57.89 & 47.54 & 26.76 \\ \hline
steinb4\_10 &50 &100 &4.00 &5.00 &5 &\bf 5 &5.00 &5 &\bf 5 &5.00 &5 &\bf 5 &5.00 &5 &\bf 5 &5.00 &5 &\bf 5 &0.00 & 0.00 & 0.00 \\ \hline
steinb4\_25 &50 &100 &4.00 &12.00 &12 &\bf 12 &12.00 &12 &\bf 12 &12.00 &12 &\bf 12 &12.00 &12 &\bf 12 &12.00 &12 &\bf 12 &0.00 & 0.00 & 0.00 \\ \hline
steinb4\_40 &50 &100 &4.00 &20.00 &20 &\bf 20 &20.00 &20 &\bf 20 &20.00 &20 &\bf 20 &20.00 &20 &\bf 20 &19.90 &19 &\bf 20 &0.00 & 0.00 & 0.00 \\ \hline
steinb10\_10 &75 &150 &4.00 &7.00 &7 &\bf 7 &7.00 &7 &\bf 7 &7.00 &7 &\bf 7 &7.00 &7 &\bf 7 &7.00 &7 &\bf 7 &0.00 & 0.00 & 0.00 \\ \hline
steinb10\_25 &75 &150 &4.00 &18.00 &18 &\bf 18 &18.00 &18 &\bf 18 &18.00 &18 &\bf 18 &17.85 &17 &\bf 18 &18.00 &18 &\bf 18 &0.00 & 0.00 & 0.00 \\ \hline
steinb10\_40 &75 &150 &4.00 &28.00 &28 &28 &27.65 &27 &\bf 29 &25.10 &24 &27 &24.35 &23 &26 &27.30 &27 &28 &7.41 & 11.54 & 3.57 \\ \hline
steinb16\_10 &100 &200 &4.00 &10.00 &10 &\bf 10 &10.00 &10 &\bf 10 &10.00 &10 &\bf 10 &10.00 &10 &\bf 10 &10.00 &10 &\bf 10 &0.00 & 0.00 & 0.00 \\ \hline
steinb16\_25 &100 &200 &4.00 &25.00 &25 &\bf 25 &25.00 &25 &\bf 25 &25.00 &25 &\bf 25 &24.35 &24 &\bf 25 &25.00 &25 &\bf 25 &0.00 & 0.00 & 0.00 \\ \hline
steinb16\_40 &100 &200 &4.00 &36.12 &36 &\bf 37 &36.00 &36 &36 &33.20 &32 &34 &32.45 &32 &34 &35.95 &35 &\bf 37 &8.82 & 8.82 & 0.00 \\ \hline
steinc6\_10 &500 &1000 &4.00 &50.00 &50 &\bf 50 &50.00 &50 &\bf 50 &50.00 &50 &\bf 50 &49.10 &47 &\bf 50 &50.00 &50 &\bf 50 &0.00 & 0.00 & 0.00 \\ \hline
steinc6\_25 &500 &1000 &4.00 &125.00 &125 &\bf 125 &122.55 &121 &124 &107.50 &106 &110 &89.90 &85 &94 &104.95 &102 &108 &13.64 & 32.98 & 15.74 \\ \hline
stienc6\_40 &500 &1000 &4.00 &145.84 &144 &\bf 147 &140.40 &139 &142 &114.10 &112 &117 &109.80 &106 &117 &121.40 &119 &125 &25.64 & 25.64 & 17.60 \\ \hline
steincc11\_10 &500 &2500 &10.00 &50.00 &50 &\bf 50 &50.00 &50 &\bf 50 &50.00 &50 &\bf 50 &50.00 &50 &\bf 50 &50.00 &50 &\bf 50 &0.00 & 0.00 & 0.00 \\ \hline
steinc11\_25 &500 &2500 &10.00 &125.00 &125 &\bf 125 &125.00 &125 &\bf 125 &125.00 &125 &\bf 125 &123.30 &122 &\bf 125 &125.00 &125 &\bf 125 &0.00 & 0.00 & 0.00 \\ \hline
steinc11\_40 &500 &2500 &10.00 &200.00 &200 &\bf 200 &200.00 &200 &\bf 200 &200.00 &200 &\bf 200 &194.25 &190 &198 &200.00 &200 &\bf 200 &0.00 & 1.01 & 0.00 \\ \hline
steinc16\_10 &500 &12500 &50.00 &50.00 &50 &\bf 50 &50.00 &50 &\bf 50 &50.00 &50 &\bf 50 &50.00 &50 &\bf 50 &50.00 &50 &\bf 50 &0.00 & 0.00 & 0.00 \\ \hline
steinc16\_25 &500 &12500 &50.00 &- &- &- &125 &125 &\bf 125 &125.00 &125 &\bf 125 &125.00 &125 &\bf 125 &125.00 &125 &\bf 125 &0.00 & 0.00 & 0.00 \\ \hline
steinc16\_40 &500 &12500 &50.00 &- &- &- &200 &200 &\bf 200 &200.00 &200 &\bf 200 &200.00 &200 &\bf 200 &200.00 &200 &\bf 200 &0.00 & 0.00 & 0.00 \\ \hline
plan50\_10 &50 &135 &5.40 &5.00 &5 &\bf 5 &5.00 &5 &\bf 5 &5.00 &5 &\bf 5 &5.00 &5 &\bf 5 &5.00 &5 &\bf 5 &0.00 & 0.00 & 0.00 \\ \hline
plan50\_25 &50 &135 &5.40 &12.00 &12 &\bf 12 &12.00 &12 &\bf 12 &12.00 &12 &\bf 12 &12.00 &12 &\bf 12 &12.00 &12 &\bf 12 &0.00 & 0.00 & 0.00 \\ \hline
plan50\_40 &50 &135 &5.40 &20.00 &20 &\bf 20 &20.00 &20 &\bf 20 &20.00 &20 &\bf 20 &20.00 &20 &\bf 20 &19.90 &19 &\bf 20 &0.00 & 0.00 & 0.00 \\ \hline
plan100\_10 &100 &285 &5.70 &10.00 &10 &\bf 10 &10.00 &10 &\bf 10 &10.00 &10 &\bf 10 &10.00 &10 &\bf 10 &10.00 &10 &\bf 10 &0.00 & 0.00 & 0.00 \\ \hline
plan100\_25 &100 &285 &5.70 &25.00 &25 &\bf 25 &25.00 &25 &\bf 25 &25.00 &25 &\bf 25 &25.00 &25 &\bf 25 &25.00 &25 &\bf 25 &0.00 & 0.00 & 0.00 \\ \hline
plan100\_40 &100 &285 &5.70 &37.00 &37 &37 &37.05 &37 &\bf 38 &35.80 &35 &37 &34.00 &33 &36 &36.00 &35 &37 &2.70 & 5.56 & 2.70 \\ \hline
plan200\_10 &200 &583 &5.83 &20.00 &20 &\bf 20 &20.00 &20 &\bf 20 &20.00 &20 &\bf 20 &20.00 &20 &\bf 20 &20.00 &20 &\bf 20 &0.00 & 0.00 & 0.00 \\ \hline
pan200\_25 &200 &583 &5.83 &- &- &- &48.95 &48 &\bf 50 &46.50 &46 &48 &41.80 &39 &43 &45.95 &45 &48 &4.17 & 16.28 & 4.17 \\ \hline
plan200\_40 &200 &583 &5.83 &- &- &- &60.65 &58 &\bf 62 &52.95 &52 &56 &49.35 &47 &51 &55.70 &54 &58 &10.71 & 21.57 & 6.90 \\ \hline
plan500\_10 &500 &1477 &5.91 &50.00 &50 &\bf 50 &50.00 &50 &\bf 50 &50.00 &50 &\bf 50 &44.95 &42 &47 &50.00 &50 &\bf 50 &0.00 & 6.38 & 0.00 \\ \hline
plan500\_25 &500 &1477 &5.91 &- &- &- &92.29 &90 &\bf 94 &78.15 &76 &80 &60.95 &57 &65 &78.20 &77 &80 &17.50 & 44.62 & 17.50 \\ \hline
plan500\_40 &500 &1477 &5.91 &- &- &- &122.31 &119 &\bf 124 &92.60 &90 &95 &82.85 &78 &86 &100.15 &97 &102 &30.53 & 44.19 & 21.57 \\ \hline

\end{tabular}}
\caption{Message-passing and multi-start greedy performances. Columns 1-4 give the characteristics of the benchmark. For each algorithm, columns 1-3 represent the average, the minimum and the max number of accommodated paths over 20 runs of a given set of commodity instance respectively. ACO and LS performances are reported in \cite{blesa04,pham2012}. Performance comparison between MP and the other algorithms is given in the three last columns, representing the performance ratio $100\cdot(M_{acc}^{BP}/M_{acc}^{alg}-1)$ where $alg$ indicates the algorithm used (MSG, ACO and LS respectively). We use as $M_{acc}^{MP}$ the best one between $MP$ with and without reinforcement.}
\label{table:benchmark}
\end{table*}

\end{document}